# Real-time imaging of quasiparticle dynamics at a topological defect in an electronic crystal

Yevhenii Vaskivskyi[1,2]*, Jaka Vodeb[1,3,4], Igor Vaskivskyi[1,2,4], Dragan Mihailovic[1,2,4]*

[1]*Department of Complex Matter, Jozef Stefan Institute, Ljubljana Slovenia.*
[2]*Faculty of Mathematics and Physics, University of Ljubljana, Ljubljana, Slovenia.*
[3]*Jülich Supercomputing Centre, Institute for Advanced Simulation, Forschungszentrum Jülich, Jülich, Germany.*
[4]*CENN Nanocenter, Ljubljana, Slovenia.*

**Corresponding authors: yevhenii.vaskivskyi@ijs.si, dragan.mihailovic@ijs.si*

**Defects formed during nonequilibrium self-assembly of quantum matter can dominate its emergent properties. Their macroscopic manifestations are typically characterised as noise, but their internal dynamics remain largely experimentally inaccessible. Making significant advances into the investigation of the microscopic degrees of freedom of such defects, we use fast scanning tunnelling microscopy to resolve, in real time, both internal and global dynamics of a mesoscopic Y-junction defect in an electronic crystal created through self-assembly after a local electromagnetic perturbation. We directly track individual electron rearrangements on millisecond timescales and map spatially localised telegraph noise characteristic of a two-level system. The phase and amplitude of these fluctuations are correlated with the observed charged particle trajectories, revealing a direct connection between collective order-parameter dynamics and microscopic charge motion. We model the dynamics as arising from the interplay of local Coulomb correlations, non-local configurational constraints and hybridised amplitude–phase collective modes bound to the junction. These constraints, together with the non-trivial broken symmetries of the defect, protect long-lived local quasiparticle configurations against external perturbations. Our results establish fast scanning tunnelling microscopy as a means of probing the internal dynamics of metastable quantum defects and reveal how microscopic correlations and collective modes are intertwined within topologically non-trivial structures in electronic crystals.**

The choreography of self-assembly in nonequilibrium systems is governed by symmetry, topology and causality constraints[1]. Systems emerging through a phase transition may form metastable spatially localised states that break long-range order, effectively forming defects. When such defects are topologically non-trivial, they are robust against external influence and consequently can be very difficult to manipulate or remove. Such situations arise anywhere from rapid crystal growth to molecular self-assembly in biological systems and in quantum many-body systems in the process of formation of emergent orders after a phase transition[2,3]. This problem extends to processes of practical importance, such as thin film deposition and crystal growth on substrates, where such defects play an important role in limiting the performance of devices. In quantum devices such as superconducting qubits, they may trap quasiparticles, forming two-level systems with internal degrees of freedom that effectively generate noise. Being able to identify and characterise such defects in detail is thus of both fundamental and practical importance.

Beyond defects formed by chance, we may engineer the formation of topologically and symmetry-protected mesoscopic metastable states with new and desired emergent properties by guiding the trajectory of the nonequilibrium system with appropriate perturbations, such as ambient temperature gradients or local charge currents. The resulting states are of interest when they show robustness against external perturbations, and particularly if they harbour dynamics on an energy scale which is below the thermal energy scale[4,5]. Electronic crystals are an excellent playground for experimentally investigating such behaviour. They controllably form metastable domains and topological defects associated with domain wall junctions in the electronic superlattice after rapid self-assembly[6–11]. Robust topological defects -solitons have previously been statically imaged by scanning-tunnelling microscopy (STM) in quasi-1D systems[12]. In 2D, the topological structure of such states is more complex, but it was shown that it can be controlled and manipulated by external perturbation[13], while domain walls have been shown to contain low-energy states[10,11,14–16] and as yet unidentified sources of noise[17,18].

Here we first create mesoscopic defects in an electronic crystal superlattice using a localised transient electrical perturbation with an STM tip. We then choose a particularly interesting metastable topologically non-trivial (Y-junction) defect at the intersection of three different domain walls in the hexagonal electronic superlattice. Using a fast-scanning tunnelling microscope (FSTM)[19–21], we record movies of density of states fluctuations in *real time* that correspond to motion of *individual polarons* – electrons surrounded by a local lattice deformation – localised within the Y-junction defect. 2D maps of tunnelling current temporal fluctuations reveal localised emergent two-level system (TLS) 'telegraph' noise whose analysis leads to a very detailed picture of the mesoscopic emergent quasiparticle states and poses fundamental questions on the origin of its slow dynamics and unprecedented robustness to external perturbations.

The electronic crystal chosen for this investigation is 1T-$TaS_2$, a layered transition-metal dichalcogenide (Fig. 1a) which exhibits multiple charge ordered phases, including a 'hidden' metastable state with intricate mesoscopic structure created by external perturbation[7,11,15,22]. Its ordered equilibrium phases have traditionally been described in terms of classical field theory with a complex charge-density-wave (CDW) order parameter. This framework captures the amplitude, phase, commensurability and discommensuration structure of the electronic modulation[23,24]. The ground state is a commensurate superlattice with two possible orientations $\chi_L$ and $\chi_R$ relative to the crystal lattice, related by mirror symmetry $\sigma$[25,26]. Microscopically, however, 1T-$TaS_2$ exhibits a propensity for electron localisation and the formation of polarons with well-defined lattice distortions, as shown in Fig. 1a(ii). A particularly relevant recent experiment has revealed the existence of individual polarons at temperatures well above any charge ordering temperature (> 600 K), and elucidated their local structural distortions[27]. The charge ordering of the system through different symmetry-breaking transitions can thus also be understood in terms of crystallisation of polarons subject to lattice strain and Coulomb repulsion[27–31], which was verified by X-ray local structure analysis[27]. The importance of the crystallisation approach is that it can help understand phenomena which traditional approaches cannot explain, such as the existence of the polaron jamming into an amorphous polaronic state[32] and metastable states exhibiting topologically arrested defect annihilation dynamics[13]. This motivates a microscopic description in which the localised polarons self-organise to form a correlated electronic crystal[31–33]. A basic Mott state description[29,34–36] may be useful for describing single layers of 1T-$TaS_2$[35,37], but is not suitable for addressing the amorphous polaronic state[32] or the structure and dynamics of vertex-antivertex annihilation in its metastable domain state[8,13]. However, the related charged Coulomb lattice gas (CLG) model of correlated polarons[31] leads to useful predictions of a variety of nonequilibrium states that have since been experimentally observed[38].

This complexity of the system makes it particularly interesting from the point of view of defect formation, and here we will find it necessary to discuss both a correlated-polaron approach for the microscopic configurations and rearrangements within the defect, and a conventional order parameter analysis to describe the slow coarse-grained collective dynamics of three intersecting long-range ordered domains.

To create a topologically non-trivial embedded defect in the surface layer, we apply an electrical pulse through a sharp STM tip, which locally destroys the C order[13–15] and typically forms quite diverse domain structures[8,11,13–15] with a variety of domain walls and junctions. The rapid self-assembly of polarons results in a mesoscopic network of topologically non-trivial domain walls and vertices[9,11,13] that can be manipulated by a metal tip or externally sourced current[13] (Fig. 1a iv). A Y-junction at the intersection of a particularly interesting set of domain walls sets the stage for this detailed study. Importantly for obtaining consistent results, the structure was found to be stable at low temperatures, and for tip currents below ~2 nA, allowing a detailed investigation. Fig. 1b shows a time-averaged topographic FSTM image of a Y-junction vertex forming at the intersection of three domains (labelled A, B and C) within such a structure. A schematic construction of the Y-junction is shown in Fig. 1c, showing the polaron positions (denoted as idealised hexagrams) at the domain boundary. Domains A and B both have the same chirality ($\chi_L$), while domain C has $\chi_R$ chirality. Choosing a clockwise path, A and B are related to each other via a translation vector $\boldsymbol{T}_{A\rightarrow B} = (3\boldsymbol{a} - \boldsymbol{b})$, where $\boldsymbol{a}$ and $\boldsymbol{b}$ are crystal lattice constants (Fig. 1c). B and C are related by a mirror operation $\boldsymbol{\sigma}_{\boldsymbol{a}-\boldsymbol{b}}$ about the $\boldsymbol{a} - \boldsymbol{b}$ axis followed by a translation $-\boldsymbol{b}$, while A and C are related by a combination of a mirror operation $\boldsymbol{\sigma}_{\boldsymbol{a}}$ about the $\boldsymbol{a}$ axis, followed by a translation $-\boldsymbol{a}$. A schematic image of the CDW order at the vertex corresponding to the FSTM image in Fig. 1b is shown in Fig. 1d.

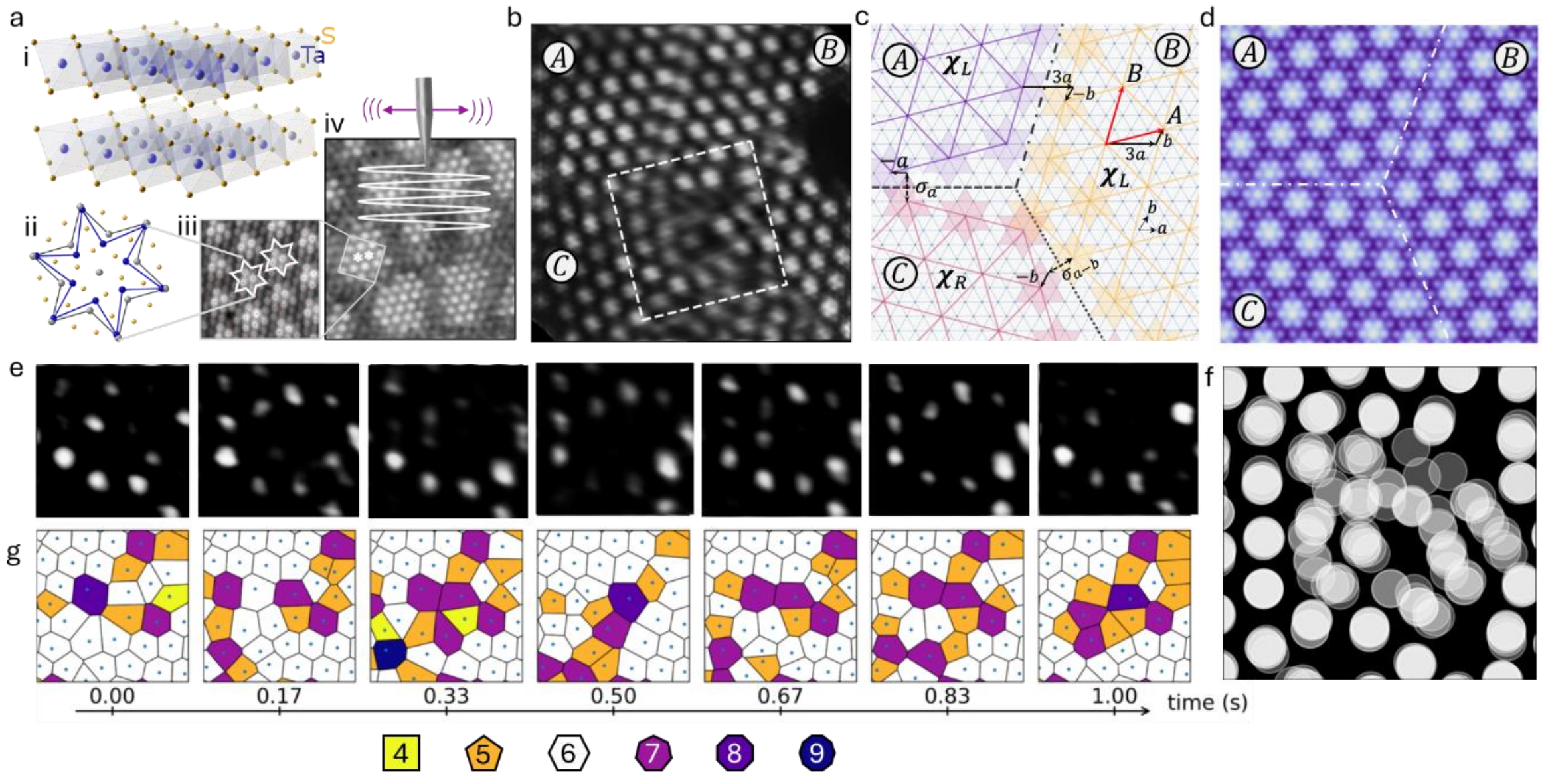

**Fig. 1. Real-time dynamics of semi-localised electrons in a vertex trap**. **a.** (i) The crystal structure of 1T-$TaS_2$, (ii) the polaronic lattice distortion at different temperatures[27] and (iii) high-resolution STM topography of the C CDW state. (iv) the domain state created by external perturbation with a schematic description of the fast-scanning STM measurement. **b.** The 7-minute time-average of an STM movie (videos S1 and S2) recorded with the STM tip at a constant height above the surface (V = 0.4 V, i = 1.5 nA) shows continuity of QP motion. **c.** The vertex structure showing broken mirror and translational symmetries, indicated by the vector constructions at the domain walls. **d.** The order parameter $\Psi_{\mathrm{CDW}}(\mathrm{x})$ at the vertex corresponding to (B) and (C) (the Ta atoms in the crystal lattice are also visible). **e.** A set of extracted frames measured 0.17 seconds apart, extracted from the full frame showing quasi-periodic individual polaron motion (see Video S3 for full sequence). **f.** A time-averaged image emphasising polaron motion. The trajectories are analysed in Fig. 3. **g.** A W-S tessellation of (E) showing 5-gon-7-gon and 4-gon-8-gon dislocations as well as single n-gon disclinations. A full movie is available in the supplement (S4).

The temporal and spatial current fluctuations $i(\boldsymbol{r}, t)$ in the vertex recorded by an FSTM reveal real-time motion of individual polarons. A movie recorded with constant tip height (voltage) over an area of $15 \times 15\ nm^2$ at a frame rate of $R_F \sim 6\ fps$ (25 MHz sampling rate, T = 77 K) is shown in videos S1 and S2. Fig. 1e shows a short sequence of frames extracted from Video S3 showing polaron movements in time on the millisecond timescale ($\tau \sim 10\ ms$). The time-integrated polaron positions (depicted by circles) are shown in Fig. 1f, emphasising their trajectories. The observed strongly sub-thermal fluctuations ($kT \gg h\nu$) are persistent on the timescale of the experiment (7 minutes). Dislocations (DL) and disclinations (DC) associated with moving polarons are revealed by a Wigner-Seitz (W-S) tessellation (Fig. 1g) and appear as pentagon/heptagon pairs or individual $N - gons$ $(N = 4,5,7,8)$. Such defects are important, as they are topologically protected, which can lead to arrested electron dynamics[39,40]. The full Video S4 of defect dynamics shows that DCs occur primarily at the intersection of the $\chi_L/\chi_R$ domains and along B-C domain walls (DWs). DLs are present at the $\boldsymbol{T}$-symmetry-breaking DW boundary and show quite lively dynamics on the millisecond timescale (see Video S4).

Time-series plots of the temporal and spatial fluctuations of the tip current $i(\boldsymbol{r}, t)$ for a few chosen positions $\boldsymbol{r}$ identified by coloured dots in Fig. 2d are shown in Fig. 2a. (A larger data set is presented in the SI). We clearly identify two types of behaviour: (i) TLS telegraph noise in regions where particle motion is observed, and (ii) 'single-level' noise far from these regions. The corresponding current distribution $\theta(i(\boldsymbol{r}, t))$ is shown in Fig. 2b. Further, we observe in Fig. 2a that the telegraph noise of trace 3 has *opposite phase* compared with traces 1, 2 and 4. Thus, we note that the overall TLS behaviour is spatially correlated and phase-locked: the effective phase relation between the pixels $\phi(t)$ is maintained throughout the duration of the measurement even for pixels that are far apart (see SI). A map of $\phi(\boldsymbol{r})$ in Fig. 2f shows phase modulations across the entire image, with reference to the phase of the centre of the vertex.

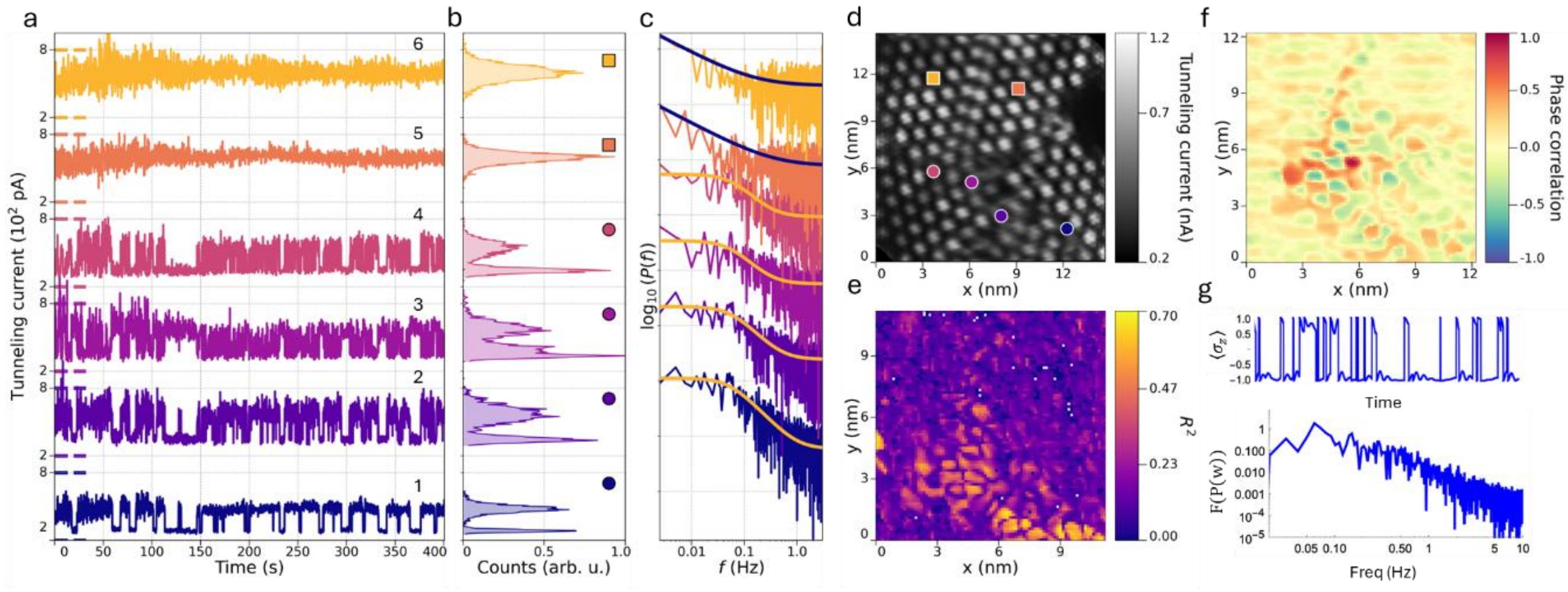

**Fig. 2. Noise analysis of charge fluctuations**. **a.** Time-traces of tunnelling current $\mathrm{i}(\mathbf{r}, \mathrm{t})$ recorded at constant tip height at a few selected points shown in **(d)**. They show temporally correlated 'telegraph noise' traces (1-4) with 0 or π phase shift $\phi$, or featureless noise (traces 5 and 6). **b.** The dwell time distributions for the traces shown in (A) showing either single level or TLS behaviour. **c.** Fits to the power spectra of $\mathrm{i}(\mathbf{r}, \mathrm{t})$ at points shown in (A) give either flat-top Lorentz spectra with $1/f^2$ tails or 1/f noise with frequency-independent tails at high f. **e.** A pixel map of the amplitude of TLS noise expressed as the fit confidence factor $\mathrm{r}^2$ of Lorentz fits to $\mathrm{F}(\mathrm{i}(\mathbf{r}, \mathrm{t}))$ shown in **(d)**. **f.** A map of the variation of phase $\phi(\mathbf{r})$ across the vertex space. **g.** Level occupancy $\langle \sigma^z \rangle$ and power spectrum F(P(w)) for an external noise-driven model TLS system.

Fourier power spectra $F_t(i(\boldsymbol{r}, t))$ corresponding to the temporal current fluctuations $i(\boldsymbol{r}, t)$ are shown in Fig. 2c. In the regions that show telegraph noise $F_t(i(\boldsymbol{r}, t))$ clearly show characteristic 'flat-top'

Lorentzian curves $L(f)$ (Fig. 2c) with a $1/f^2$ high-frequency tail. Elsewhere, we ubiquitously see $1/f$ power spectra - attributed to current originating from many TLSs with different level spacings[17,41–43]. These areas show no saturation at low frequencies and exhibit frequency-independent 'white' noise at high frequencies, presumed to be of thermal origin. Separating the three different noise contributions $S = L(f) + \beta/f + \gamma$ for each pixel, we can spatially separate TLS noise from $1/f$ noise and obtain high-resolution maps of the amplitude of the TLS noise source (Fig. 2f). It is relevant to note here that the FSTM tip frame rate (6 Hz) is much faster than the observed polaron TLS dynamics shown in Fig. 2. We also note that there is no external source of noise in the observed TLS frequency range, that could drive the dynamics directly.

In addition to the amplitude and phase fluctuations of $i(\boldsymbol{r}, t)$ above, plotting the polaron *centre positions* in the $x - y$ plane as a function of time (Fig. 3a and Video S5), we see that some particles appear to show dumbbell-like oscillating motion (Fig. 3b). The polaron motion patterns are also clearly discernible in the FSTM movies (S1 and S2). To examine if and how these displacements are correlated, in Fig. 3c and 3d we present an analysis of the positional correlations of pairs of polarons. The coloured dots depict the correlations between pairs of adjacent polarons along the *x* and y coordinates (see SI for details). We observe particularly non-trivial polaron motion at the vertex core. For example, an abrupt change from symmetric (red dots) to antisymmetric polaron-pair displacements (blue dots) is visible at the centre of the vertex. Further away from the vertex, the displacements along the $\boldsymbol{T}$-symmetry-breaking A-B DW boundary show antisymmetric motion, seen as a line of pale blue dots along the A-B domain wall.

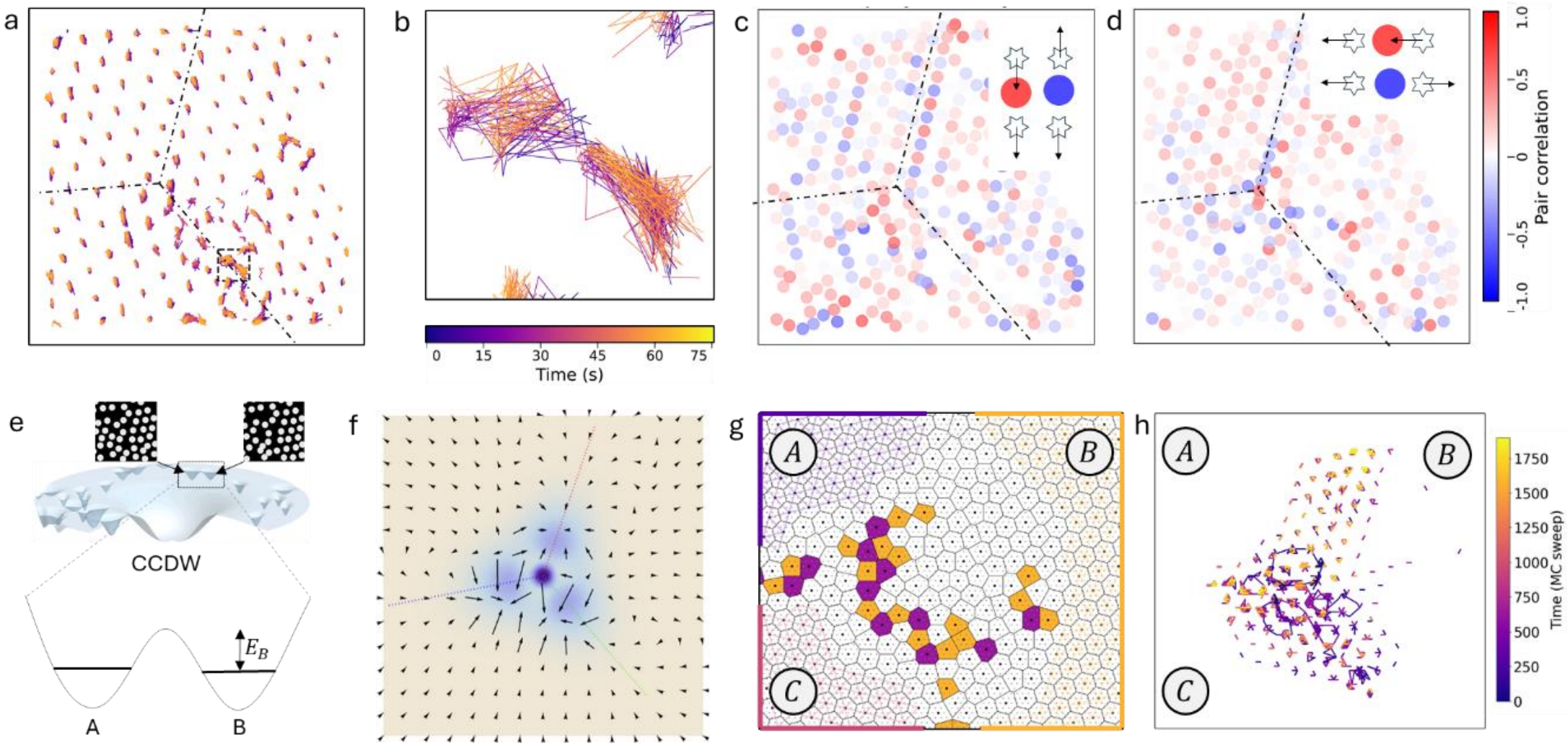

**Fig. 3. Inter-particle motion correlations and modelling of Y-junction dynamics**. **a.** and **b.** the measured spatial trajectory of polarons. The timeline is indicated by the colour bar. **c.** The polaron-pair displacement correlation coefficient $\xi$ (indicated by colour) along the x direction and **d.** The y direction. The displacements are defined by the legend. Note the cross-over from symmetric to antisymmetric motion at the vertex core. **e.** An energy landscape describing the domain configurational space. Neighbouring configurational states described by a double-well potential giving rise to TLS behaviour shown in Fig. 2g. **f.** A model simulation showing $\Delta^2(\boldsymbol{r}, t)$ (blue colour) and vector field proportional to the gradient $\nabla\phi(\boldsymbol{r}, t)$ for the low-energy branch of the hybridised Higgs-Goldstone mode. (See Video S6 for the temporal fluctuations. **g.** An MC simulation of defect time evolution. See Video S8 for the full sequence. **h.** Local MC polaron trajectories (See also Video S9).

Turning to a theoretical understanding of the dynamics, it is clear that neither a macroscopic CDW theory nor a microscopic, correlated electron model on their own can account for the rich set of observations described above. In particular, the microscopic motion of individual polarons at the core of the defect is difficult to model in terms of CDW theory but can be faithfully reproduced using correlated polaron dynamics. In contrast, an order parameter approach is useful in understanding the overall slow dynamics at the Y junction in terms of a low-frequency Goldstone collective mode. Since collective modes are supported both by correlated polaronic crystals as well as CDWs, we don't run into contradictions. We thus proceed with a theoretical discussion on two levels. A correlated-polaron model is used to describe the microscopic re-configuration dynamics subject to Coulomb interactions between individual polarons, which are also subject to constraints imposed by the surrounding domain topology.
The collective order parameter, understood here as either the coarse-grained order parameter of the polaron crystal or directly describing a CDW, describes the amplitude and phase fluctuations that drive transitions between these re-configurations.

We first consider the microscopic polaron dynamics. The polarons are described by the Hamiltonian (19):

$$H = -\sum_{i,j} t_{ij} c_i^{+} c_j + \sum_{i<j} V_{i,j} n_i n_j \qquad (1)$$

where $t_{i,j}$ is the polaron tunnelling amplitude between sites $i$ and $j$, $c_i^{+}, c_i, n_i$ are the spinless fermion creation, annihilation, and occupation number operators at site $i$, $V_{i,j} = \frac{V_0}{r_{ij}} exp\left(-\frac{r_{ij}}{r_s}\right)$ is the Coulomb screening potential, $r_{ij}$ is the real space distance between sites $i$ and $j$, $r_s = 4.5\ a \sim 1.5\ nm$ is the screening radius, $V_0 = \frac{e^2}{\epsilon_0 a} \sim 1\ eV$, and $a$ is the lattice constant we can calculate the polaron dynamics at the Y junction. Previous work has shown that domain-wall networks can host fractionalised charge and topologically entangled configurations that strongly constrain local polaron motion (*12*). For the Monte Carlo calculation, we consider the classical interaction-dominated limit $t_{ij} = 0$ and impose the experimentally observed domains as fixed boundary conditions around the model space. These non-local constraints restrict the allowed polaron rearrangements at the Y-junction. The simulated defect evolution, shown as a Wigner–Seitz tessellation in Fig. 3g and Video S8, tends to form dislocation chains along the two mirror-symmetry-breaking domain walls. The corresponding trajectories in Fig. 3h and Video S9 qualitatively reproduce the confined and directionally correlated polaron motion observed experimentally in Fig. 3a, demonstrating the role of non-local boundary constraints in the microscopic dynamics.

We next consider the collective fluctuations that drive transitions between the allowed polaron configurations. At larger length scales, the system is described by the conventional complex order parameter commonly written as $\Psi(\boldsymbol{r},t) = \Delta(\boldsymbol{r},t) e^{i\phi(r,t)}$, where $\Delta(\boldsymbol{r},t)$ is the "Higgs" amplitude and $\phi$ describes the phason or "Goldstone mode". In commensurate systems such as 1T-$TaS_2$, the bulk Goldstone mode is gapped with an energy scale corresponding to the pinning energy [41]. Approaching the Y junction, $\Psi(\boldsymbol{r},t)$ is suppressed as $\Delta(\boldsymbol{r},t) \to 0$ and the *local fluctuations* of the phase stiffness $K(\boldsymbol{r},t) \propto \Delta(\boldsymbol{r},t)^2$ slow down and diverge in relative magnitude as $\frac{\delta K(r,t)}{K} \sim \frac{\delta \Delta(r,t)}{\Delta}$ . Near the Y-junction, the spatially varying order parameter and broken translational and mirror symmetries couple the amplitude and phase fluctuations, giving rise to hybridised collective modes. To describe these dynamics we use a phenomenological Ginzburg–Landau Lagrangian for the two collective fields, $\Delta(\boldsymbol{r},t)$ and $\phi(\boldsymbol{r},t)$,

$$\mathcal{L} = \frac{\chi}{2}(\partial_t \phi)^2 + \frac{1}{2v^2}(\partial_t \Delta)^2 - \frac{\rho_s}{2}(\nabla\phi)^2 - \frac{\kappa}{2}(\nabla\Delta)^2 - \frac{\mu^2}{2}(\Delta - \Delta_0)^2 - \frac{\rho_0}{2}\Delta^2(\nabla\phi)^2. \quad (2)$$

The first two terms describe the inertial dynamics of the phase and amplitude fields, respectively. The terms proportional to $\rho_s$ and $\kappa$ penalise spatial variations of the phase and amplitude, while the term proportional to $\mu^2$ restores the amplitude towards its equilibrium value $\Delta_0$. The final term makes the phase stiffness dependent on the local amplitude and therefore couples the Higgs and phase degrees of freedom. In a uniform system the two modes are approximately distinct, whereas the strongly inhomogeneous order parameter near the Y-junction leads to their hybridisation. The corresponding equations of motion are obtained from the Euler–Lagrange equations for $\Delta$ and $\phi$. (See SI for calculation details). We are particularly interested in the low-frequency hybrid branch localised at the Y vertex core to describe the observed dynamics. The predicted fluctuations of the phase for this mode are primarily longitudinal along the DWs, but also include skew motion, arising from the non-trivial broken symmetries at the Y junction, as shown in Fig. 3f. The phase currents, which are proportional to the charge currents – and drive polaron motion, are given by $\boldsymbol{j}_\phi(\boldsymbol{r},t) = -K(\boldsymbol{r},t)\nabla\phi(\boldsymbol{r},t)$, depicted by the arrows in Fig. 3f. (Videos S6 and S7 show the temporal fluctuations of $\boldsymbol{j}_\phi(\boldsymbol{r},t)$.) The observed TLS behaviour can be thought to arise from transitions between neighbouring minima of the constrained polaron configurational landscape (Fig. 3e). Each minimum represents a locally stable polaron arrangement compatible with the Y-junction boundary conditions. Collective amplitude–phase fluctuations modulate the barriers between these configurations, producing telegraph switching and the corresponding flat-top Lorentzian noise spectrum shown in Fig. 2g. Thus, the microscopic model determines which polaron rearrangements are allowed, while the collective mode drives transitions between them. A calculated noise spectrum $S_k(\boldsymbol{r},\omega) = \int dt\, e^{i\omega t}\langle\delta K(\boldsymbol{r},t)\delta K(\boldsymbol{r},0)\rangle$, calculated using Langevin dynamics is enhanced and shifted to low frequencies, with the appearance of telegraph-like noise and a flat-top spectrum (Fig. 2g), like the experimental spectra in Fig. 2c (See SI for calculation details).

**Discussion**

The observed polaron dynamics at the Y-junction defect are consistent with transitions between discrete, correlated configurations, rather than the free motion of isolated electrons (or polarons). These transitions are governed by local Coulomb interactions and constrained by the non-local topology of the surrounding domains, while the slow time scale of dynamics is consistent with collective amplitude-phase fluctuations. The persistence of this motion under continuous STM measurement underscores the robustness of this topologically constrained degree of freedom against local perturbations. While electrical noise of unidentified origin has been observed in diverse systems — such as 1T-$TaS_2$[17,41], Al-Si superconducting circuits[45], and strongly correlated materials[46,47], where it is even associated with local pairing[48] — our FSTM mapping pioneers the identification, characterisation, and potential control of such two-level systems (TLSs) in metastable quantum states. This approach could be extended to other systems, including glasses, other quantum materials with fabrication-induced defects, or purposefully engineered defects for quantum circuits.

The domain structures and junctions observed here are not unique to 1T-$TaS_2$ but are ubiquitous in transition metal chalcogenides[31,49,50] and other systems with similar phase diagrams. For instance, discommensurations in the correlated-polaron phase diagram suggest that deviations from magic-number filling fractions ($f = 1/13, 1/9, 1/7$)[31] lead to domains and junctions of varying sizes and shapes. This has direct relevance to doped pseudogap phases[50], where incompletely ordered phases may host various defects. Furthermore, hexagonal lattices are not a prerequisite: square lattices of misfit dislocations also exhibit arrested dynamics at domain wall crossings[51], hosting local excitations. A particularly striking parallel is the ultraslow-frequency noise spectrum observed in transmon qubits[52], suggesting that the mechanisms uncovered here may contribute to understanding and mitigating noise in quantum computing devices.

# Methods

## Materials

Single-crystal 1T-$TaS_2$ samples used in the research were in-house grown using the chemical vapour deposition method, as described previously[7,26].

## Fast STM

The fast-scanning electronics module (Elettra-Sincrotrone Trieste S.C.p.A.) is implemented on top of an LT Nanoprobe STM system (Scienta Omicron AB). The system records the tunnelling current $i_t(V)$ from the Pt/Ir tip to the surface of the 1T-$TaS_2$ crystal with a tip that controllably oscillates laterally across the sample surface (Fig. 1a). The measured STM tunnelling current at time $t$ and position $\boldsymbol{r}$ is $i(\boldsymbol{r},t) \propto |M_{tip-sample}|^2 D_s(\boldsymbol{r},t)$ , where $M_{tip-sample}$ is the usual matrix element for the tunnelling transition between the tip states and the sample, $D_s(\boldsymbol{r},t)$ is the density of states at the surface of the sample at any given time $t$. The total current $i(\boldsymbol{r},t)$ is an incoherent integration over time of the individual electron tunnelling currents as the tip moves across the sample. The $V-i$ curves in conventional scanning mode are consistent with those reported previously[9,11,14,15]. The $V$ and $i$ parameters are shown in the figure captions for each case. In the C state, the bright 'blobs' approximately correspond to individual polarons. In the vertex region and in the domain walls, the CDW acquires non-trivial time-dependent phase fluctuations $\boldsymbol{\delta}(\boldsymbol{r},t)$, so this approximation is no longer valid, and $D_s(\boldsymbol{r},t)$ is determined by the structure of microscopic electronic wavefunctions[9,11,14,15]. The Fast STM typically records $100 \times 100$ pixel image sequences with a tunnelling current sampling rate of $25\sim100$ MHz, averaged typically for $10\sim20$ $\mu s$ per pixel limited by the amplifier bandwidth (400 kHz), recorded line-wise (Fig. 1b), resulting in frame rates $R_F = 5\sim100$ Hz. Note that all dynamics that is slower than the individual frame recording time $t > \tau_{frame}(\sim0.1\ s)$ is visible as real-time motion. Dynamics that occurs on timescales $\tau_{frame} > t > \tau_{pixel}(\sim10^{-8}s)$ leads to temporal and spatial distortions of the images. Faster dynamics with $t \ll \tau_{pixel}$ averages out and the structure appears static. If there is no motion blur, the FTSM gives atomic-resolution images.

## Monte Carlo simulations of the defect dynamics at a T-junction

Monte Carlo simulations of the Markov chain dynamics for the Hamiltonian on a triangular lattice were performed using the standard Metropolis algorithm. The proposed successive state in the chain consists of setting $n_i = 0$ at an occupied site and $n_j = 1$ at a randomly selected neighbouring unoccupied site. The simulations were performed with a fixed number of charged particles on a lattice of the size of $80 \times 80$ atomic sites with lattice period $a$. The boundary conditions were fixed in the sense that for approximately 2 rows of polarons from the edge, polarons cannot move. This imposes the ordering of polarons towards the middle of the system, resulting in a highly frustrated topologically imposed structure in the exact same way as in the experiment. The number of polarons was chosen to be 492, closely resembling the experiment. The interaction potential $V(i,j)$ in the simulations shown in the paper has a Yukawa form $V_{i,j} = \frac{V_0}{r_{ij}} exp\left(-\frac{r_{ij}}{r_s}\right)$ with the screening radius $r_s = 4.5\ a$ and was cut at distances larger than $24a$. Note that the screening length is larger than the polaron radius in the C state. The annealing simulations were performed by decreasing the temperature from $0.005V_0$ to $0.003V_0$, which is deep within the ordered phase, with 2000 temperature steps with equal increments, where 70 Monte Carlo sweeps ($70 \times 80 \times 80$ steps) were performed at each temperature point. The initial configuration was chosen by setting all the polarons that are able to move next to each other in the centre of the system.

## Data availability

All the data supporting the results of this study are available within the Article and Supplementary Materials. Additional data are available from the corresponding author upon request.

## Acknowledgements

We wish to acknowledge discussions with Viktor Kabanov, Lev Vidmar, Leo Radzihovski, Alexei Kitaev, Duncan Haldane, Mikhail Feigelman, Anton Zeilinger, Michel Devoret, Rok Zitko, Denis Golež and Serguei Brazovskii. A large language model was used as a Mathematica programming aid for part of the simulations shown in Fig. 3f and Fig. 2g and the Supplement. All the generated code parts were manually checked.

We want to acknowledge funding which supported the research presented in the Article. Slovenian Research and Innovation Agency (ARIS) grants P1-0040 (YV, IV, DM), N1-0290 (IV, DM), and I0-0053 (DM). European Research Council (ERC) AdG project HIMMS (DM).

## Author contribution

YV, IV and DM conceived the research project. YV and IV performed STM measurements. JV performed Monte Carlo simulations. DM performed calculations for the hybridised modes. YV, JV and DM performed data analysis and visualisation of the results. DM wrote the original manuscript. All the authors contributed to the discussion of the results, manuscript editing and writing of the supplementary materials.

# Supplementary Information for

### Real-time imaging of quasiparticle dynamics at a topological defect in an electronic crystal

Yevhenii Vaskivskyi[1,2*], Jaka Vodeb[1,3,4], Igor Vaskivskyi[1,2,4], Dragan Mihailovic[1,2,4*]

[1] *Department of Complex Matter, Jozef Stefan Institute, Ljubljana, Slovenia.*
[2] *Faculty of Mathematics and Physics, University of Ljubljana, Ljubljana, Slovenia.*
[3] *Jülich Supercomputing Centre, Institute for Advanced Simulation, Forschungszentrum Jülich, Jülich, Germany.*
[4] *CENN Nanocenter, Ljubljana, Slovenia.*

**Corresponding authors: yevhenii.vaskivskyi@ijs.si, dragan.mihailovic@ijs.si*

Experimental data acquisition

The STM data is measured using a combination of `LT Nanoprobe` (Scienta Omicron AB, Sweden) multi-tip scanning tunnelling microscope and a `FAST SPM` (Elettra Sincrotrone Trieste S.C.p.A., Italy) module. The experiments discussed in the main text were performed at 77 K with liquid nitrogen as a coolant in the bath cryostat of the STM.

The image rasterisation is performed by applying a sine wave along one axis (X', fast axis) and a triangular wave along the second axis (Y', slow axis) as shown in Fig. S1. The image size and framerate are chosen to be compatible with the mechanical properties of the piezo drives controlling the tunnelling tip at this temperature, as well as their physical arrangement in the instrument (as shown in the Figure).

The Fast STM is performed in the quasi-constant-height mode, with the feedback loop being much slower than the tip modulation along the fast axis. To compensate for the sample orientation relative to the scanning axes (tilt), additional modulation is performed in the Z' axis synchronously to the X' and Y' axes. The experiment's feedback setpoint (stabilised tunnelling current) is kept at the level of 1.0 to 1.5 nA to ensure both a good signal-to-noise ratio and low electrical disturbance to the sample in case of the conventional STM measurement.

The scanning parameters for the movie presented in the main text are: 100 × 100 px with a pixel frequency of 119990 Hz, which corresponds to a 6 frames per second output movie (refer to the next section for the pixel data conversion details).

Experimental data conversion

The tunnelling current is recorded as a 1D array of values with a 16-bit ADC. To reconstruct the fast movie, the following steps are performed:

1. The current time trace is cut into frames.
   A fast scan of a size $X_{px} \times Y_{px}$ and length $N_f$ frames is represented by an array of size $2 \times \mathrm{x_{px}} \times \mathrm{y_{px}} \times N_f$ due to the "subframes" representing the "forward" and "backward" directions (half-periods of the harmonic driving signal, shown in Fig. S1b-c).
2. The fast grid of a corresponding size is triangulated onto the regular square grid using the standard Delaunay triangulation method.
   The fast grid represents equally spaced points in time. Due to the shape of the harmonic signal, this would correspond to the spatial distribution as shown in Fig. S1. The current values are interpolated linearly for the regular grid. The resolution used in the experiment ensures an average of 8 pixels per CDW unit cell to coincide with the resolution usually used for such studies.

3. The drift correction is done on the frames (Fig. S2).
   The drift between each pair of frames is found independently in $X_r$ and $Y_r$ (orthogonal) directions using standard methods. The overall drift of the scanning tip relative to the sample is found as a square function for each direction (Fig. S2a).
   The found drift is used to redo Step 2. The perfect fast grid is offset by the corresponding drift value before the triangulation to the regular square grid is performed. This allows us to avoid double interpolation to obtain the drift-corrected images (Fig. S2b).

These three steps ensure as little data processing as possible before it is used in further analysis. At the same time, for plotting and visualising purposes in the Main text and the Supplementary Information, additional contrast adjustment or data filtering can be performed on the images to highlight features or parts of interest (Fig. S2c).

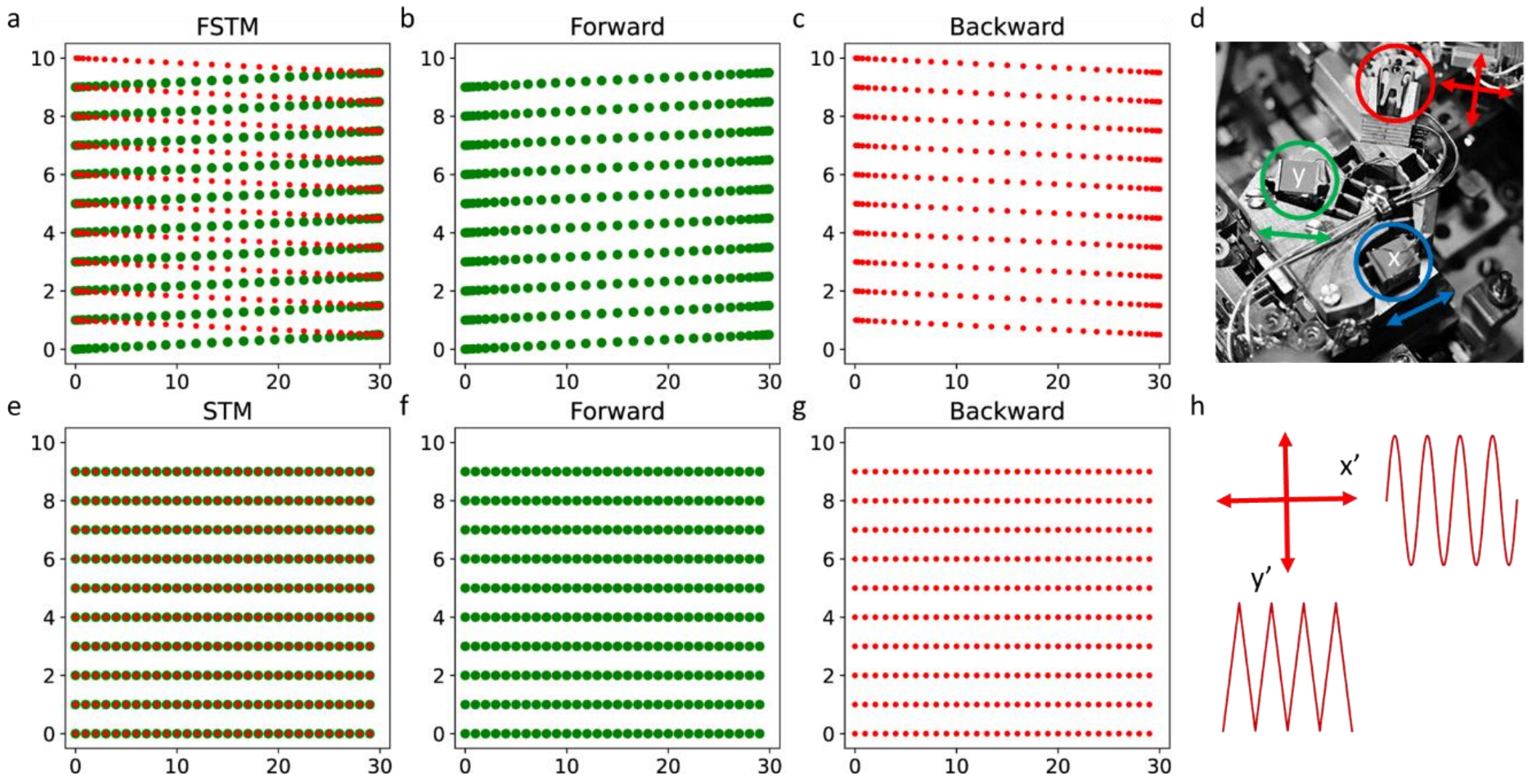

**Fig. S1.** Fast STM raster compared to the conventional STM. **a.** Fast STM raster of a single "frame" and its components (**b**, **c**). **e.** Conventional STM raster and its components (**f**, **g**). **d.** STM scanner photo with the tip (red) and two piezo elements (green and blue) marked. Note that the relative arrangement of the actual device X and Y piezo drives is not orthogonal. **h.** Temporal shape of the voltage signal applied to x' and y' piezo.

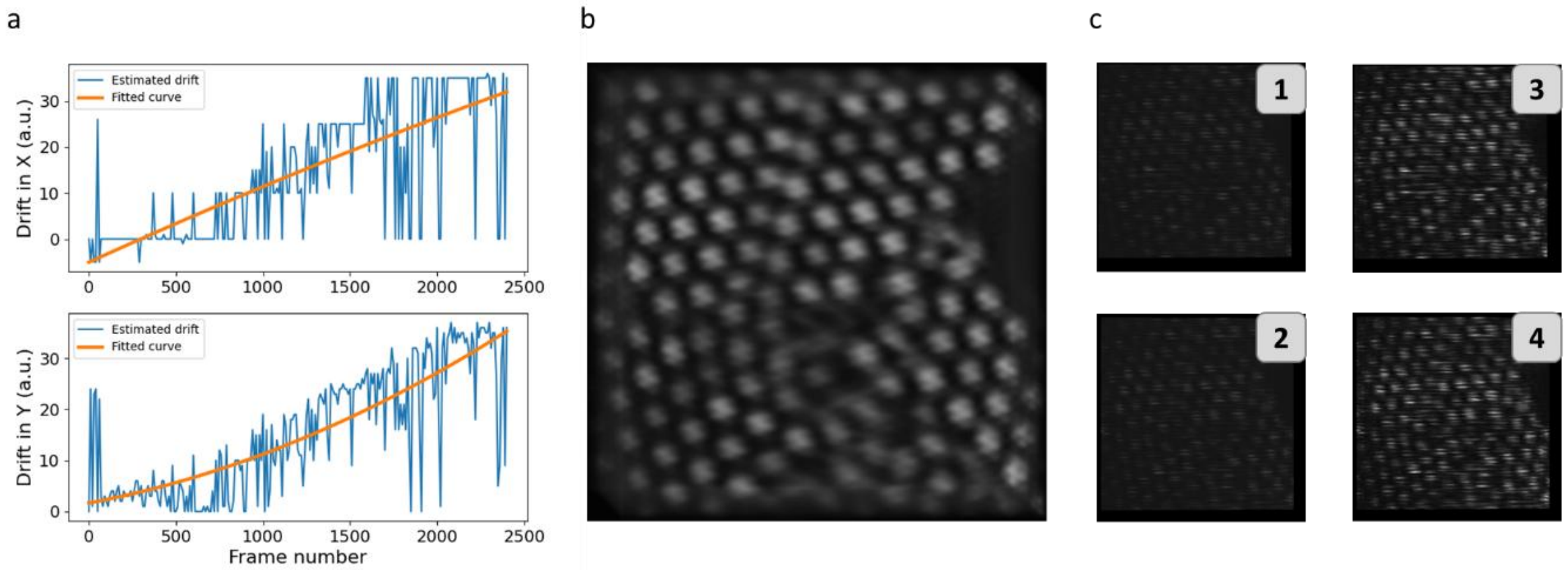

**Fig. S2.** Drift compensation for the fast STM movie. **a.** Example of the drift between the frames for a time sequence. **b.** The sum of all the frames showing the "average" frame over the movie. This panel shows the different domains of the CDW in 1T-$TaS_2$, which stay static for all the frames and the dynamic part in between, which is getting blurred when taking the sum. **c.** Example of raw frames after the drift compensation (1, 2) and after the visualisation adjustment (3, 4 – same frames as 1, 2) – in this case, only outliers by the current value are masked (0.01 and 99.99 percentiles).

<u>Single- and multi-level current signals</u>

If we consider time traces of the measured tunnelling current at some positions on the surface, we can clearly see two types of behaviour: single-level-like and two-level-like (telegraph noise) as shown in Fig. S3a. All the positions for this analysis are defined from the polaron positions defined for the reference (time-averaged) image of the surface (Fig. 1b of the main text). The first case is observed for the positions inside the domains (curve 1 in the figure), while the second is characteristic of the area of "dynamic" polaron behaviour (curves 2-6). Points with the telegraph noise show different ratios between the "high" and "low" states, with only a single one (curve 7) showing a clear preferential level. At the same time, for several positions (curves 8-10), three different levels can be defined. Most of the dynamic points show some kind of temporal synchronisation, which can clearly be seen at times 110-145 s, 360-370 s and others. At the same time, they can have opposite phases as shown in the section "Signal phase analysis".

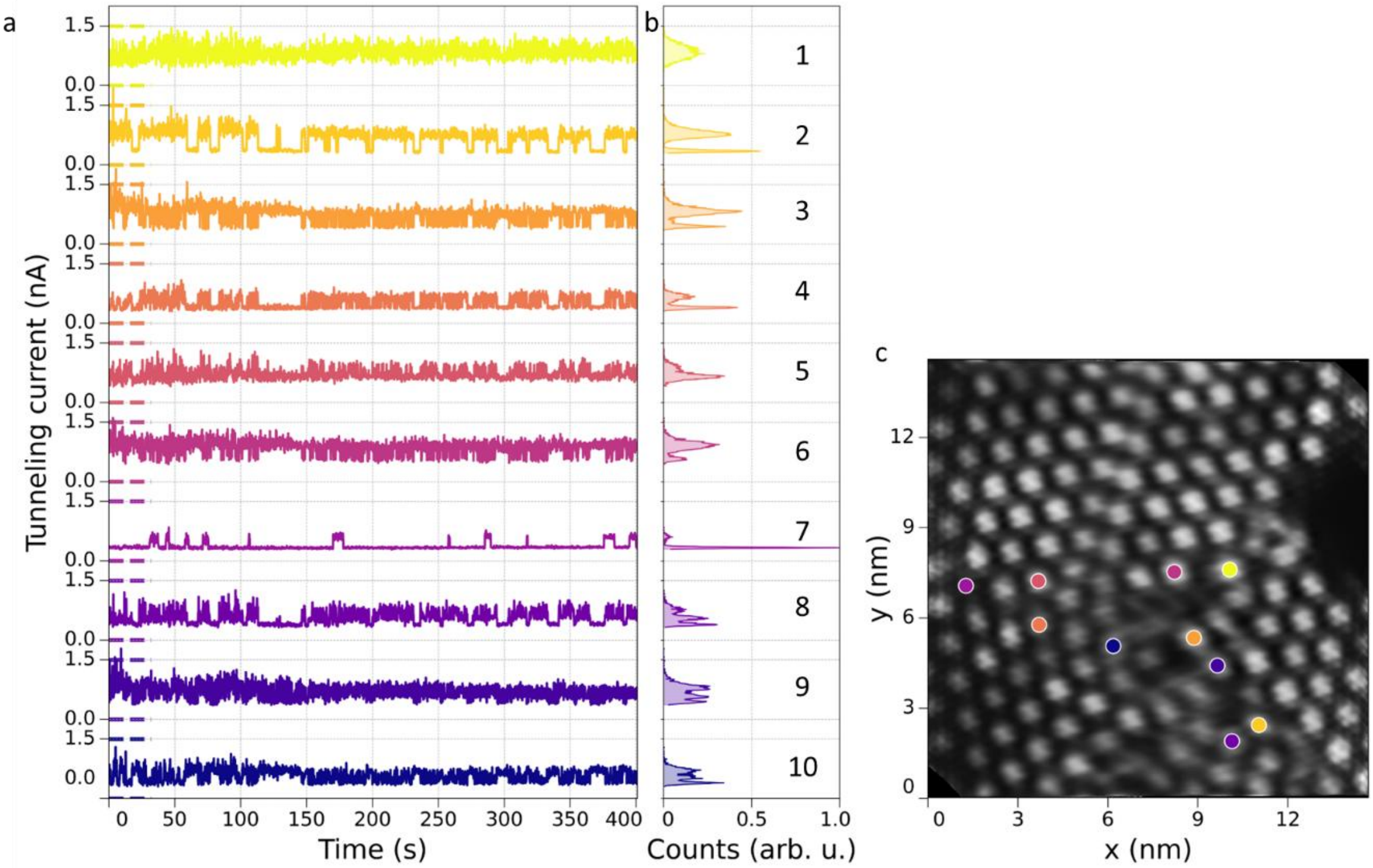

**Fig. S3.** Examples of the single- and multi-level behaviour of the current signal measured for the selected polaron positions. **a.** Time traces of the current value measured at the selected positions centred on the "polaron positions". **b.** Current values distribution for the traces from (**a**). **c.** Reference frame with the positions marked and colour-coded in the same way as in (**a**) and (**b**).

<u>Current level noise</u>

As can be seen in the previous section and from the main text, the observed "width" of the current measured seems to depend on its absolute value, which is most prominent in the case of two-level behaviour (dynamic polarons), where the "low" level looks sharper than the "high" level. To check whether this is a signature of the energetic levels' properties on the material surface or an instrumental limitation, we performed the following analysis:

1. We took all the points with the single-level behaviour of the current:

    a) The first case corresponded to the regular square grid over the entire surface (Fig. S4c).
    b) The second case corresponded only to the polaron positions as defined before (Fig. S4d).

2. We fitted a standard Gaussian function to the noise (current) distribution for each spatial point, considering and analysing the distribution of the peak FWHM from the peak centre position.
3. We fitted a linear function to the distribution from (2).

As can be seen from Fig. S4, except for some outliers, most of the extracted data points fit well to the linear dependence between the measured current and the noise distribution in the measurement. This implies that the observed effect of the level width distribution is mostly instrumental.

While the analysis was performed only on the "static" points, we consider this data as representative of the whole dataset since the measured current values cover a wide range of current values. The multi-level points were excluded due to the additional uncertainty added by the multi-peak Gaussian fitting.

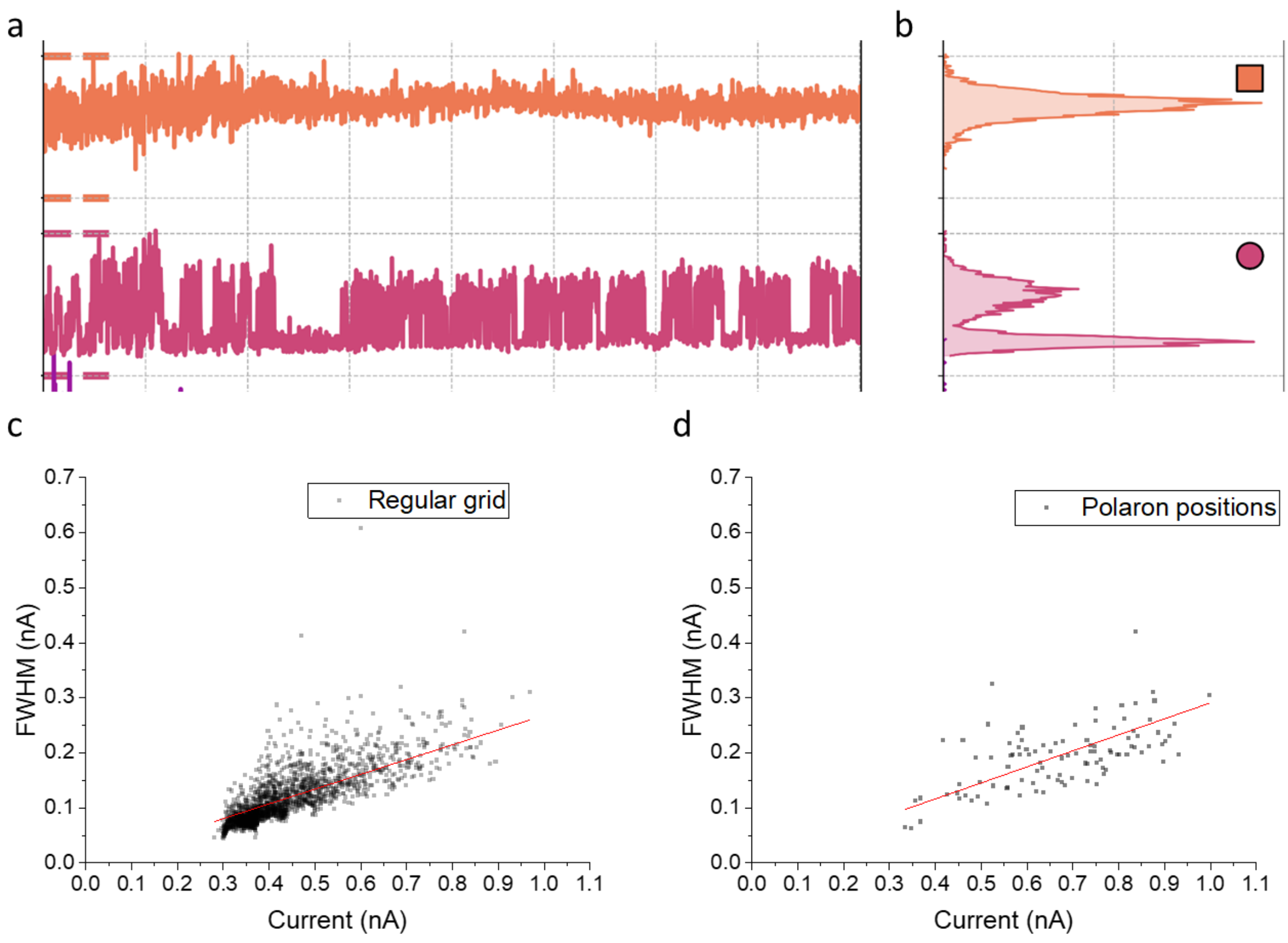

**Fig. S4.** Dependence of the noise on the absolute current level measured for a single sample position. **a.** Time trace of the tunnelling current measured for single static (one level) and dynamic (two levels) points (cut from Fig. 2 or the main text) and **b.** The corresponding current distribution. **c, d.** Dependence of the FWHM of a Gaussian peak fitted to the current level (only single-level cases) from its centre position for the regular grid over the whole scanned area (**c**) and only on the polarons positions (**d**) as defined in the main text and Fig. 1.

Frequency-domain analysis

The frequency-domain (noise) analysis is performed on the drift-corrected raw data as discussed in Section `Experimental Data Conversion`. The analysis of each spatial point is done in the following way:

1. The time trace of a single point is found as $I_{(x,y,r_s)}$, where $x, y$ are the coordinates and $r_s$ – is the size of the window in which the mean intensity value is calculated for each point in time. The window size for analysis is constant for analysis of all the points of a single movie and allows for compensating for single-pixel noise in the data, if any. For computational ease and to avoid extra conversion operations, the window is taken as a square of size $(2r_s + 1) \times (2r_s + 1)$, centred on the position $(x, y)$.
2. The frequency-domain behaviour of the point data is calculated using a standard Fast Fourier Transform (FFT).
3. The frequency data fitting.
   The power spectrum ($P(f)$) is fitted with the following functions to find the type of signal present at the point:

   a. Lorentzian function corresponding to the single oscillator behaviour:
   $$F(f) \approx a\frac{1}{f^2 + \gamma^2} + c$$
   For this fitting, the Lorentzian is considered centred at $x_0 = 0$.
   b. $1/f$ function corresponding to the average of many oscillators:
   $$F(f) \approx b\frac{1}{f} + c$$

   After the fits are found, they are compared in terms of the following characteristic values:

   i. Coefficient of Determination (how well a model explains the variance in the data)
   $$R^2 = 1 - \frac{\Sigma(f_i - \hat{f}_i)^2}{\Sigma(f_i - \overline{f}_i)^2}$$
   Higher value $R^2$ is better;
   ii. Root Mean Square Error
   $$RMSE = \sqrt{\frac{1}{n}\sum(f_i - \hat{f}_i)^2}$$
   Lower value is better.

4. The best model is found as the one with higher $R^2$ and lower $RMSE$.
5. Movie map reconstruction.

For the full movie, the frame is split into a regular "downsampled" grid, and for each point, Steps 3 and 4 are performed, resulting in a spatial map of the frequency features (Fig. S5).

To ensure that the statistics (and as a result, the fitting) is the same, the edges of the movie frame obtained during the drift correction process are not considered in the map.

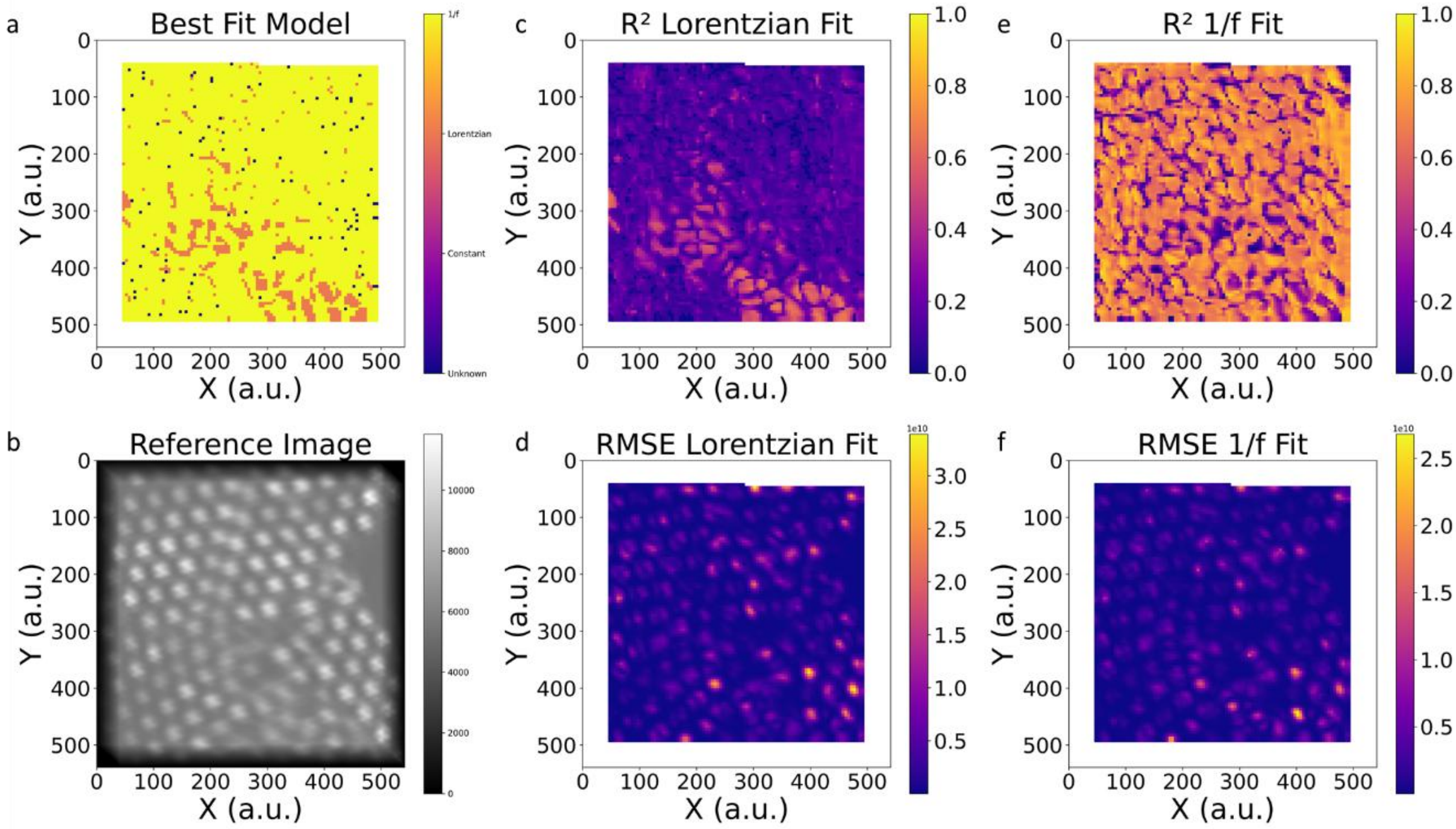

**Fig. S5.** Spatial map of the sample features. **a.** Best fit model distribution obtained as the highest $R^2$ and the lowest $RMSE$ for the selected spatial point on the surface as referenced in panel **(b)**. Spatial maps of $R^2$ and $RMSE$ are shown for the Lorentzian fit **(c, d)** and $1/f$ fit **(e, f)** correspondingly.

<u>Signal phase analysis</u>

Since the sample shows two-level-like behaviour as seen from Fig. 2 of the main text, we can analyse the phase component of such a signal across the surface of the sample. For this, we can take a single pre-defined point and calculate the relative phase of the rest of the movie in the following way.

1. As a reference, we take a time-dependent signal of a single position on the sample (e.g., corresponding to a defined polaron position). The time trace is calculated for a square window of pixels, as discussed in Section `Frequency-domain Analysis`. This signal is taken as $I_{ref}$.
2. For a point with coordinates $(x, y)$, we calculate its time trace $I_{(x,y)}$ in a similar fashion.
3. We calculate a *sliding-window* Pearson correlation coefficient $r_{k_{(x,y)}}$:
$$r_{k_{(x,y)}} = \frac{\sum\left(I_{(x,y)_i} - \bar{I}_{(x,y)_i}\right)\left(I_{ref_i} - \bar{I}_{ref_i}\right)}{\sqrt{\sum\left(I_{(x,y)_i} - \bar{I}_{(x,y)_i}\right)^2 \sum\left(I_{ref_i} - \bar{I}_{ref_i}\right)^2}}$$
Two examples of such correlation traces are shown in Fig. S6d.
4. A single point value of the phase is defined as a mean value of the correlation trace:
$$\mathrm{r}_{(x,y)} = \overline{r_{k_{(x,y)}}}$$
5. The map is reconstructed for a regular grid.

Several examples of the signal phase calculation are shown in Fig. S6a-c. As can be seen, the features of the map do not depend on the selected reference within the same phase. When a point of the opposite phase is selected as the reference, the map inverts (Fig. S6e). For a static position on the sample with no two-level-like behaviour, no features are observed in the phase map (Fig. S6f).

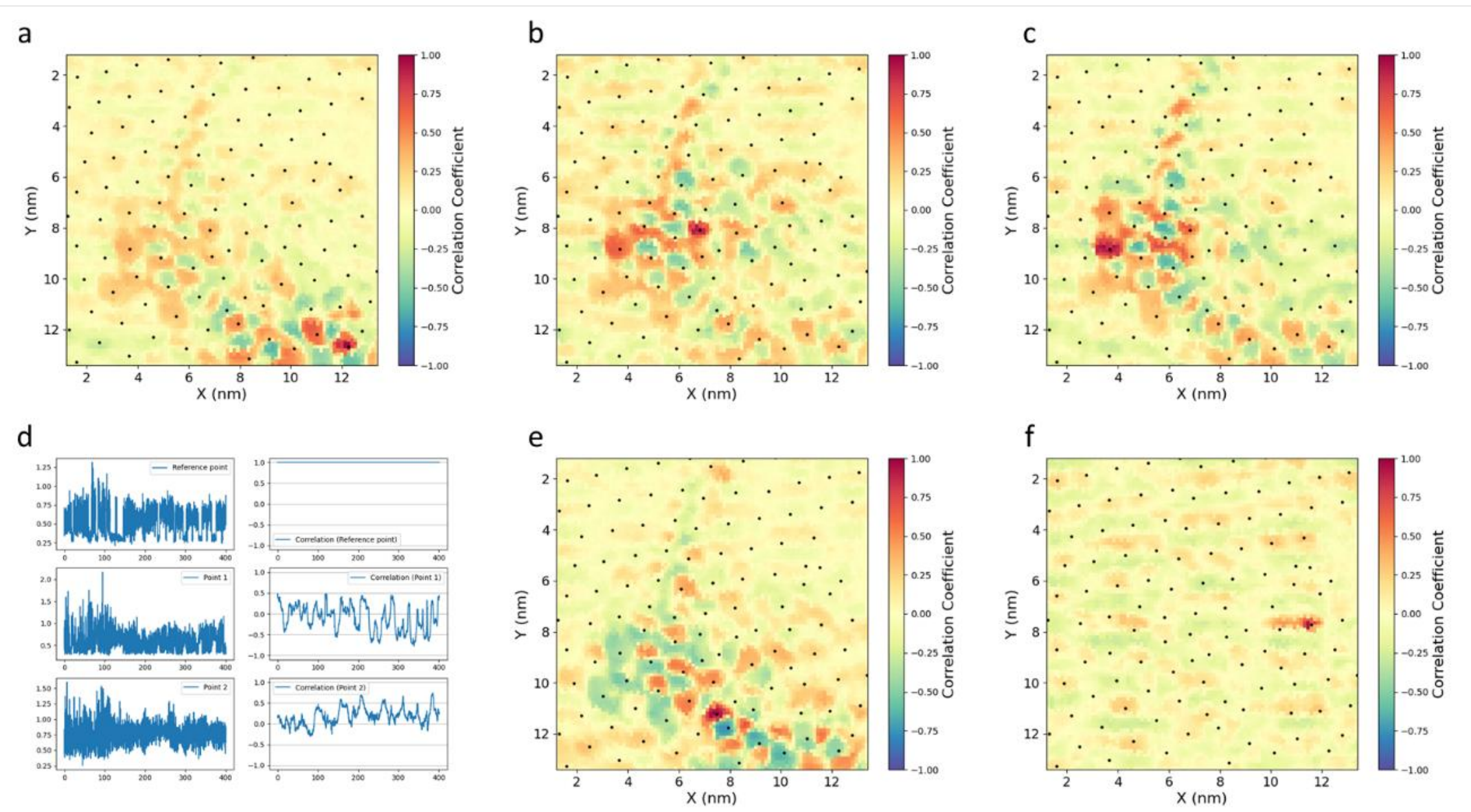

**Fig. S6.** Signal phase analysis. **a.** The phase map clearly demonstrates areas with a two-level-like signal. Positive and negative values correspond to the in-sync and anti-sync signals. The maximum value of 1 (red) corresponds to the reference point. **b, c.** Same as panel **(a)** for different reference points of the same phase. **d.** Examples of the single-point correlation traces ($r_{k_{(x,y)}}$) for the reference point, a two-level-like point of the opposite phase, and a static position on the sample (right) and the corresponding time traces (left). **e.** The anti-sync to **(a-c)** point taken as a reference inverts the map. **f)** The signal phase map calculated by taking a static point as a reference shows no features.

<u>Pair correlation for polaron positions</u>

The pair correlation analysis shown in Fig. 3c of the main text is done in the following way:

1. A dataset is taken in the form of polarons' positions on the surface across the movie.
2. A global (common for all polarons) quadratic drift trend is removed to avoid offset in the correlation calculation in the later steps.
3. For all the known polarons, a Voronoi reconstruction is done to find nearest neighbours (NNs).
4. For each pair of NNs, a Pearson correlation coefficient is calculated using the formula from the previous section for the full time trace of X and Y coordinates independently.
5. The corresponding coefficient from (4) is plotted as a point centred at the mid-point between the two Voronoi sites.

The corresponding results for X and Y coordinates are shown in Fig. S7. The same results for X are also presented in Fig. 3c of the main text. It can be seen in the figure that the Y coordinate results show a periodic (every second) change in the sign of the correlation coefficient for the top of the movie. This can be attributed to the experimental noise in the movie, where frames scanned "up" and "down" have different Y intensity distributions for the top and bottom of the polarons observed.

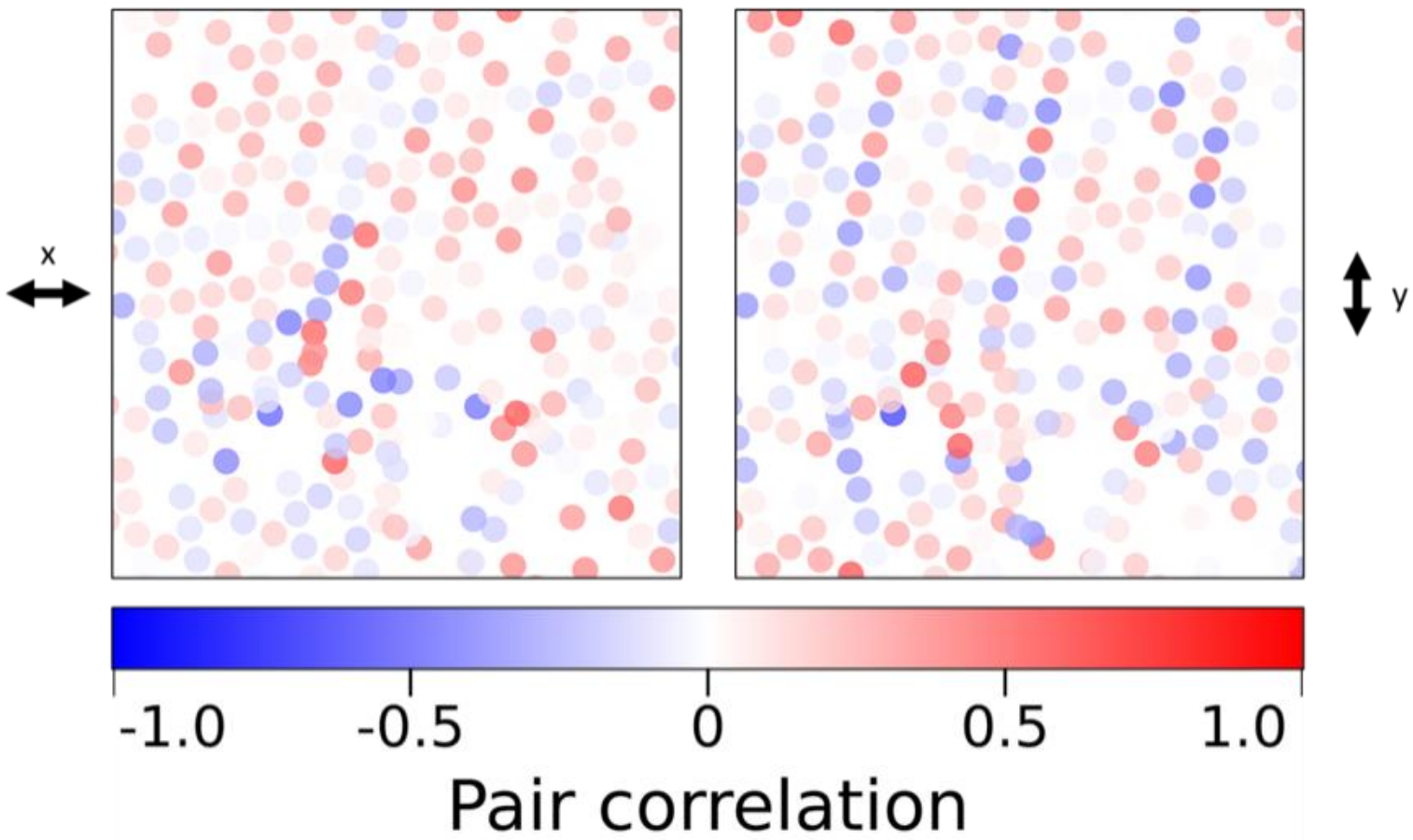

**Fig. S7.** Pair correlations for the positions of the nearest neighbour polarons in X and Y directions.

Properties of the $\boldsymbol{\sigma} - \boldsymbol{T}$ vertex in 1T-TaS2

The structure of the vertex shown in the movie and Fig. 1 of the main text is shown in Fig. S8. The figure corresponds to the experimentally observed structure. The three superlattice domains A, B and C are individually fit to the time-averaged topographic movie image, taking into account drift and motional averaging. The crystal lattice (fine triangular mesh) is superimposed without further adjustment of the domain positions. The orientation of the David stars remains the same, as it is fixed by the crystal lattice.

Because the $\boldsymbol{Q}_i$ are restricted by the lattice commensurability condition, the net phase circulating around the vertex $\oint_l \nabla\phi . d\boldsymbol{l}$, where $\phi = \sum_{i=1}^{3}(\boldsymbol{Q}_i . \boldsymbol{r} + \varphi)$, does *not* in general result in a winding number of 0 or $2\pi n$, where $n$ is an integer. In particular, the combination of mirror and translation operation at the AB wall, as well as additional translations at BC and AC domain walls, introduces non-trivial phase jumps. The phase jump across the mirror wall $\psi_B(r) = \psi_A(\sigma\, \boldsymbol{r})^*$, (* is complex conjugation caused by M) is not simply additive, as it reverses the sign of the imaginary part of the order parameter $\psi_i = A_i e^{i\phi} \rightarrow \psi_i^m = A_i e^{-i\phi}$.

The transformation operators representing the domain walls AB, BC and CA, respectively, are described below.

(i) Domains A and B are related by a translation vector $\boldsymbol{T}_{A\rightarrow B} = (3\boldsymbol{a} - \boldsymbol{b})$, such that $\boldsymbol{T}_{A\rightarrow B}\chi_L^B = \chi_R^A$. Thus, the order parameter $\psi_B(\boldsymbol{r}) = \psi_A(\boldsymbol{r} + \boldsymbol{T}_{A\rightarrow B})$, etc.

(ii) For clockwise winding domains B and C are related by a mirror operation $\sigma_{\boldsymbol{a}-\boldsymbol{b}}$ about the $\boldsymbol{a} - \boldsymbol{b}$ axis, followed by a translation $-\boldsymbol{b}$ along the $\boldsymbol{a}$ axis $M_{B\rightarrow C}(\sigma_{a-b}, -\boldsymbol{b})\chi_L^B = \chi_R^C$ . For anticlockwise winding $M_{C\rightarrow B}(\sigma_{a-b}, -\boldsymbol{a})\chi_R^C = \chi_L^B$ . The order parameter change is $\psi_C(\boldsymbol{r}) = \psi_B(\sigma\, \boldsymbol{r} + \boldsymbol{a})^*$ for clockwise winding, and $\psi_C(\boldsymbol{r}) = \psi_B(\sigma\, \boldsymbol{r} - \boldsymbol{a})^*$ for anticlockwise winding.

(iii) Domains A and C are related by a combination of a mirror operation $\sigma_{\boldsymbol{a}}$ about the $\boldsymbol{a}$ axis, followed by translation -$\boldsymbol{a}$ axis, written as: $M_{A\rightarrow B}(\sigma_a, -\boldsymbol{a})\chi_L^A = \chi_R^C$ for clockwise, and $M_{B\rightarrow A}(\sigma_a, \boldsymbol{a})\chi_R^C = \chi_L^A$ for anticlockwise winding. The order parameter change is $\psi_B(\boldsymbol{r}) = \psi_A(\sigma\, \boldsymbol{r} - \boldsymbol{a})^*$ for clockwise winding, and $\psi_C(\boldsymbol{r}) = \psi_B(\sigma\, \boldsymbol{r} + \boldsymbol{a})^*$ for anticlockwise winding.

In summary, global chirality is broken by the mirror walls. The net winding is non-zero due to the phase and chirality mismatch, resulting in a vortex-like defect which may attract charge and give rise to localised in-gap states as well as pinning of collective excitations (e.g. phase modes of the CDW). Under non-equilibrium conditions, chiral CDW currents are possible around such a defect.

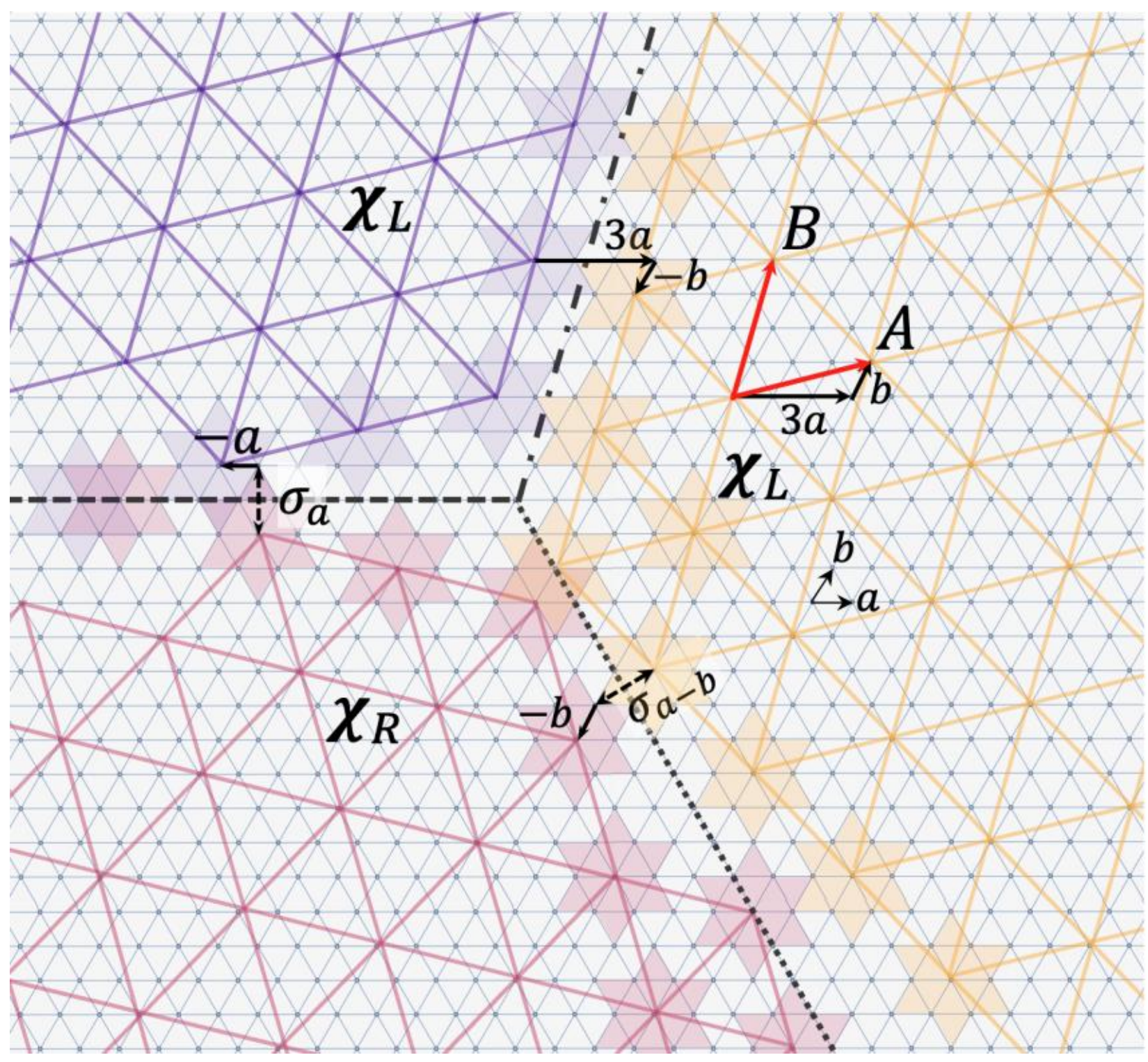

**Fig. S8.** Symmetry-breaking properties of the M-T vertex. The legend shows the crystal lattice and superlattice vectors, respectively. Domains A and B have the same chirality with respect to the crystal axes ($\chi_L$). They are related to each other through a translation $T_{A-B}$. Domain C is of opposite chirality ($\chi_R$). In addition to the mirror symmetry breaking at the A-B and B-C domain walls, there is an additional translation in both cases. The legends show lattice vectors ***a,b***, the superlattice vectors ***A, B***, and the mirror ($\sigma$) and translation (***T***) operations shown by dashed and solid arrows, respectively.

Hybrid Goldstone–Higgs modes in the 1/13 commensurate CDW superlattice

**Coupled amplitude–phase dynamics**

The complex CDW order parameter is written as

$$\Psi(\mathbf{r},t) = \Delta(\mathbf{r},t)\, e^{i\phi(\mathbf{r},t)},$$

with amplitude $\Delta$ and phase $\phi$ describing, respectively, the "Higgs" and "Goldstone" sectors of the condensate.

At the DW, and particularly at the Y junction $\Delta(\mathbf{r},t) \rightarrow 0$. Furthermore, due to breaking of symmetry at the DW and Y-junction, the two components of the order parameter can mix, leading to hybridisation. In the commensurate $1/13$ charge–density–wave (CDW) phase of *1T*–$TaS_2$, the electronic density modulation forms a $\sqrt{13} \times \sqrt{13}$ superstructure. The underlying lattice vectors **a** and **b** of the hexagonal layer define the real–space basis, while the commensurate superlattice is generated by

$$\mathbf{A} = 3\mathbf{a} + \mathbf{b}, \qquad \mathbf{B} = -4\mathbf{b} + 3\mathbf{a}.$$

The corresponding reciprocal vectors $\mathbf{G}_A$ and $\mathbf{G}_B$ are defined by $\mathbf{G}_A \cdot \mathbf{A} = 2\pi$, $\mathbf{G}_B \cdot \mathbf{B} = 2\pi$, and $\mathbf{G}_A \cdot \mathbf{B} = \mathbf{G}_B \cdot \mathbf{A} = 0$.

A translation of a CDW domain by a lattice vector **T** induces a discrete phase slip

$$\Delta\phi(\mathbf{T}) = \frac{1}{13}(\mathbf{G}_A \cdot \mathbf{T} + \mathbf{G}_B \cdot \mathbf{T}),$$

reflecting the $1/13$ commensurability.

The general time–dependent Ginzburg–Landau functional can be written as

$$\mathcal{L}_0 = \frac{1}{2}M_\Delta \dot{A}^2 + \frac{1}{2}M_\phi \dot{\varphi}^2 - \frac{1}{2}K_\Delta A^2 - \frac{1}{2}K_\phi \varphi^2 - G\,A\varphi.$$

where $M_\Delta, M_\phi$ are effective inertial parameters, $K_\Delta, K_\phi$ are stiffnesses setting the bare Higgs- and (Goldstone-) phason-like frequencies, $G$ describes Higgs-phason hybridization arising from spatial overlap of amplitude and phase gradients at the junction.

The undamped equations of motion follow from the Euler-Lagrange equations:

$$M_\Delta \ddot{A} + K_\Delta A + G\varphi = 0,$$
$$M_\phi \ddot{\varphi} + K_\phi \varphi + GA = 0.$$

In both the Y–junction and circular geometries, the term $(\nabla\phi)^2$ couples the phase rotation (Goldstone sector) to the amplitude oscillation (Higgs sector), creating a mixed collective mode. Dissipation can be included phenomenologically in the equations of motion via a Rayleigh dissipation function

$$\mathcal{R} = \frac{1}{2}\gamma_\Delta \dot{A}^2 + \frac{1}{2}\gamma_\phi \dot{\phi}^2,$$

where $\gamma_\Delta$ and $\gamma_\phi$ are damping coefficients for the amplitude-like and phase-like components, respectively, which yields to damped equations of motion:

$$M_\Delta \ddot{A} + \gamma_\Delta \dot{A} + K_\Delta A + G\varphi = 0,$$
$$M_\phi \ddot{\varphi} + \gamma_\phi \dot{\varphi} + K_\phi \varphi + GA = 0.$$

The role of the damping is to reduce amplitude and slow down the motion.

**Y–junction of domain walls**

At a trijunction of three CDW domains, domain walls meeting at $\sim 120°$ carry distinct microscopic translations:

$$\mathbf{T}_{\mathrm{DW}} = 3\mathbf{a} - \mathbf{b}, \qquad \mathbf{T}_{\mathrm{C1}} = -\mathbf{b}, \qquad \mathbf{T}_{\mathrm{C2}} = -\mathbf{a}.$$

The first, $\mathbf{T}_{\mathrm{DW}}$, preserves mirror symmetry and mainly breaks translational invariance, whereas $\mathbf{T}_{\mathrm{C1}}$ and $\mathbf{T}_{\mathrm{C2}}$ break both mirror and translational symmetry in opposite chiral senses. The corresponding phase gradients along each arm are proportional to $\Delta\phi(\mathbf{T}_i)$ and define local longitudinal wavevectors $k_i = \Delta\phi(\mathbf{T}_i)/\ell_\phi$, where $\ell_\phi$ sets the relaxation length of the phase ramp.

Within each domain arm, the static phase profile is modeled as

$$\phi_i(\mathbf{r}) = k_i\, s_i\, w(s_i)\, g_i(n_i) + \chi_i\, \frac{n_i}{\ell_\phi}\, w(s_i)\, g_i(n_i),$$

where $s_i$ and $n_i$ are the longitudinal and transverse coordinates with respect to the $i$th arm, $w(s_i)$ is a logistic envelope confining the arm to one side of the vertex, $g_i(n_i) = e^{-n_i^2/2\sigma_i^2}$ represents transverse confinement, and $\chi_i$ encodes mirror–symmetry breaking ($\chi_{\mathrm{C1}} = -\chi_{\mathrm{C2}}$ for opposite handedness).

The full dynamic phase field is a time–dependent superposition of these arm components:

$$\phi(\mathbf{r},t) = c_T(t)\,\phi_{\rm DW}(\mathbf{r}) + c_{C1}(t)\,\phi_{\rm C1}(\mathbf{r}) + c_{C2}(t)\,\phi_{\rm C2}(\mathbf{r}) + \phi_{\rm tw}(\mathbf{r},t),$$

with slowly varying weights

$$c_i(t) = 1 + \delta_i \sin(\omega_G t + \varphi_i),$$

and a small vertex–localized "twist" field $\phi_{\rm tw}(\mathbf{r},t) \propto (n_{\rm C1} - n_{\rm C2})\, e^{-r^2/2}$ that breaks the residual threefold symmetry. The amplitude field $\Delta(\mathbf{r},t)$ has a similar envelope structure but oscillates in time as

$$\Delta(\mathbf{r},t) = \Delta_0(\mathbf{r})\,[1 + a_0\cos(\omega_H t)].$$

The coupled oscillations at the Goldstone frequency $\omega_G$ and Higgs frequency $\omega_H$ produce a local hybrid mode at the vertex in which the phase currents $\mathbf{v}_\phi = \nabla\phi$ periodically shear and rotate between the three arms. The simulation of Sec. 3 visualizes $\rho(\mathbf{r},t) = |\Psi|^2 = \Delta^2$ as a density map and $\mathbf{v}_\phi$ as arrows (Fig. 3f of main text and Supplementary Video S6).

**Circular (vortex–like) hybrid mode**

A second class of collective excitation occurs when the phase field forms a continuous circulation around a topological core. The order parameter is written in polar coordinates $(r,\theta)$ as

$$\Psi(r,\theta,t) = [\Delta_0(r) + \delta\Delta(r,t)]\exp\{i[m\theta + \eta_2(t)\cos(2\theta)\,f_2(r) + \eta_3(t)\sin(3\theta)\,f_3(r)]\},$$

where: $m = \pm 1$ is the vortex charge (direction of circulation); $f_{2,3}(r)$ are Gaussian radial envelopes confining higher angular harmonics near the core; $\eta_{2,3}(t)$ are slowly varying harmonic amplitudes, modulated at the Goldstone frequency $\omega_G$; $\delta\Delta(r,t) = \Delta_0(r)a_0\cos(\omega_H t)$ describes the breathing of the amplitude (Higgs) mode.

The functions

$$\eta_2(t) = A_2\sin(\omega_G t + \varphi_2), \qquad \eta_3(t) = A_3\cos(\omega_G t + \varphi_3),$$

introduce an angular wobble or twist of the phase fronts, causing the direction of $\mathbf{v}_\phi = \nabla\phi$ to evolve continuously in time. The simultaneous amplitude breathing and phase wobble constitute a localised *Goldstone–Higgs hybridised mode* (Supplementary Video S7).

**Energy Comparison Between the Y–Junction and Circular Hybrid Modes**

In the commensurate $1/13$ charge–density–wave (CDW) phase of 1*T*–$TaS_2$, the phase of the order parameter is strongly locked to the lattice through the commensurability potential. The energy cost of collective excitations can therefore be understood in terms of the elastic stiffness $\rho_s$, the amplitude stiffness $\kappa$, and the commensurate

***(a) Y–junction hybridised mode.***

The Y–junction consists of three domain walls meeting at a vertex, each wall characterised by a local phase slip $\Delta\phi_i = \mathbf{G}\cdot\mathbf{T}_i/13$. Because the phase variations are confined along three narrow arms, the total energy is localised and finite:

$$E_{\rm Y} \simeq 3\tau\,\ell + E_{\rm vertex},$$

where $\tau$ is the line tension (energy per unit length) of a domain wall and $\ell$ is the local relaxation length of the phase ramp along the arm. There is no net $2\pi$ phase winding around the vertex, so the Y–junction hybrid does not incur global topological cost. The mode energy therefore saturates at the vertex and typically lies at low frequency. At the intersection of three domain walls, the phase currents periodically shear between the arms, producing a hybrid mode localised at the vertex. The dynamic mixing of translational and chiral arms represents a "three–pronged" collective excitation.

***(b) Circular (vortex–like) hybridised mode.***

In contrast, the circular hybrid mode carries an azimuthal phase winding $\phi(\mathbf{r}) \sim m\theta$ with integer vorticity $m = \pm 1$. Such a configuration produces a circulating phase gradient and hence a finite supercurrent density $\mathbf{j}_\phi = \rho_s \nabla\phi$ throughout the region. Its energy in a continuous medium is approximately

$$E_{\text{vortex}} \approx \pi \rho_s m^2 \ln \frac{R}{\xi} + E_{\text{core}},$$

where $\xi$ is the core (Higgs) healing length and $R$ is the system size. In the commensurate CDW, the lock–in potential further converts the logarithmic divergence into a *linear confinement*: $E_{\text{vortex}} \propto \kappa R + E_{\text{core}}$, since the phase winding must be sustained by a closed loop of discommensurations.

Around a topological defect or discommensuration loop, the CDW phase can circulate continuously. Higher harmonic terms ($\cos 2\theta$, $\sin 3\theta$) modulate this flow in time, creating a breathing, twisting vortex. The resulting mode interpolates between pure rotational Goldstone motion and isotropic amplitude breathing.

***(c) Relative ordering.***

Because the Y–junction mode involves only localised shears and amplitude modulations along three arms, while the circular mode requires a global phase rotation, we have the hierarchy

$$E_{\text{Y}} < E_{\text{vortex}},$$

with the Y–junction hybrid mode being the *lower–energy* collective mode.

<u>Simulation of telegraph noise</u>

Telegraph noise typically refers to a two-level fluctuating signal that randomly switches between two discrete states over time. It is often associated with stochastic processes such as trapping/de-trapping of charges in semiconductors or fluctuations in quantum systems. Here, we aim to describe the current fluctuations arising from fluctuations of the CDW phase $\phi$ of the order parameter in the $TM$ intersection.

Assuming a potential energy landscape shown in Fig. S9 (also in Fig. 3 of the main text), we can model the TLS system as either quantum or classical dynamics, which are applicable in different temperature regimes. Presently, due to the relatively high temperature (77 K), below we present a calculation of the Langevin dynamics. We note that the results of a quantum Lindblad calculation are qualitatively similar.

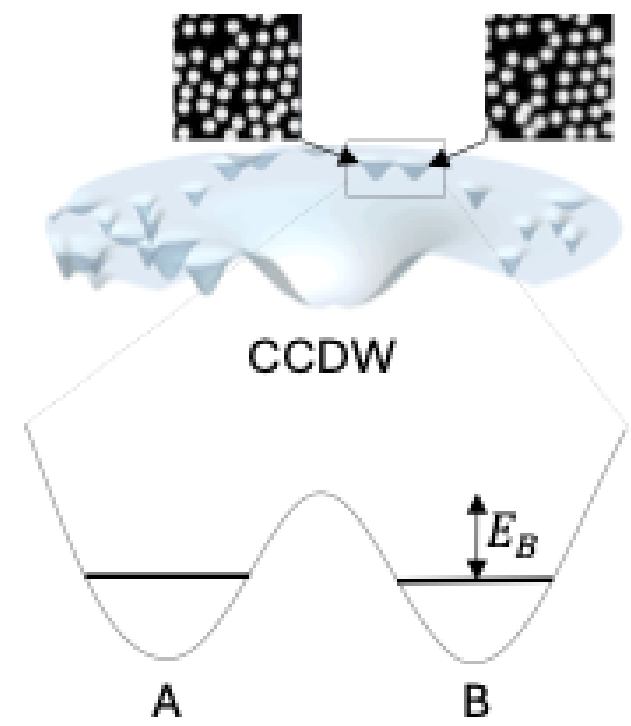

**Fig. S9.** A schematic energy landscape depicting multiple near-degenerate configurational domain states and the C ground state as the global minimum. Thermally driven and noise-driven tunnelling processes are indicated.

**Langevin TLS dynamics**

The simplest case to consider is a double-well particle potential $V(x) = ax^4 - bx^2$, (Fig. S9) where $x$ is the configurational coordinate, and a and b are constants. The appropriate equation of motion depicting dynamics driven by Brownian motion (overdamped Langevin equation) is given by:

$$\gamma \frac{dx}{dt} = -\frac{dV}{dx} + \sqrt{2\gamma k_B T}\,\eta(t) + D(t)$$

where $\gamma$ is the damping, $\eta(t)$ is the noise source and $D(t)$ – is an external drive, which is here the hybridised low-frequency collective mode.

The trajectory of the system is calculated using the Euler-Maruyama method and is shown in Fig. S10 (and Fig. 2g of the main text) for the case of a sinusoidal drive $D(t)$ that simulates the STM tip scan across the sample.

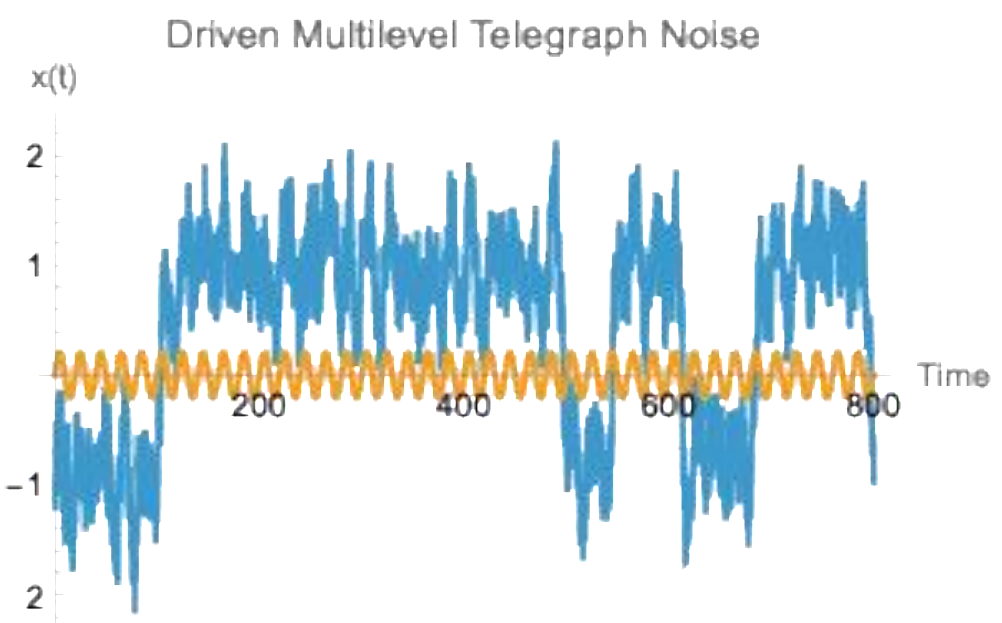

**Fig. S10.** The noise trajectory for a Langevin oscillator using a non-linear double-well potential and a sinusoidal external perturbation (orange).

The corresponding state occupancy histogram shown in Fig. S11 shows an asymmetric two-peak structure that appears if we introduce a slight asymmetry of the potential, but the peak widths are similar. Note that in the experiment, the peak *widths* are significantly different, which cannot be simulated by the present model, even by introducing different state lifetimes, multiple levels or modest changes to the potential. The power spectrum that corresponds to the trajectory in Fig. S10 is shown in Fig. S12 and Fig. 2g of the main text.

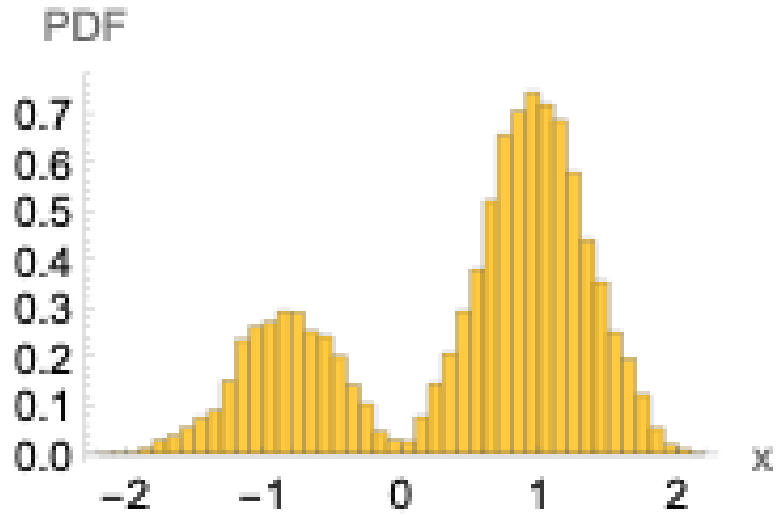

**Fig. S11.** The state occupation histogram corresponding to the calculation shown in Fig. S10. Note the asymmetric shape of the distribution.

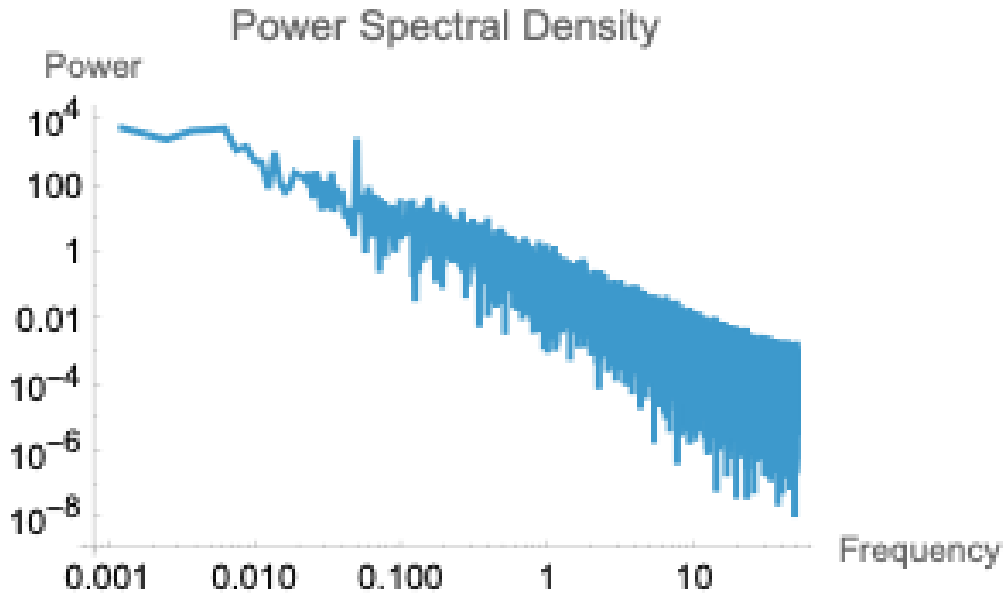

**Fig. S12.** Shows the power spectral density that corresponds to the trajectory in Fig. S10. Note the flat top response at low frequencies and $1/f^2$ behaviour at higher frequencies. The sharp peak corresponds to the sinusoidal drive frequency used in this simulation. Its presence has no significant effect on the general behaviour unless it exceeds the stochastic term $\eta(t)$.

# Supplementary Videos for

### Real-time imaging of quasiparticle dynamics at a topological defect in an electronic crystal

Yevhenii Vaskivskyi, Jaka Vodeb, Igor Vaskivskyi, Dragan Mihailovic

**Supplementary Video S1.**

https://www.youtube.com/watch?v=khGoK60tQn0&list=PLBvdEMM5_r8s&index=1

Sequence of Fast STM frames on the surface of 1T-$TaS_2$ at 77 K (6 FPS). The raw data measured with the Fast STM is shown in real time. The data is drift-corrected with no additional processing except for the outlier filter (0.05 and 99.95 percentiles), and the colour scale is adjusted for the best visualisation. The time counter represents the data time.

**Supplementary Video S2.**

https://www.youtube.com/watch?v=O190yB-rM1w&list=PLBvdEMM5_r8s&index=2

Filtered sequence of Fast STM frames on the surface of 1T-$TaS_2$ at 77 K. The Fast STM movie is shown at 5x speed, highlighting the circular motion. The data is drift corrected. Additional directional blur (vertical) and Gaussian blur (both directions) per frame is applied. The colour scale is adjusted for the best visualisation. The time counter represents the data time.

**Supplementary Video S3.**

https://www.youtube.com/watch?v=RlVYd8l6VlQ&list=PLBvdEMM5_r8s&index=3

Circular individual polaron motion on the surface of 1T-$TaS_2$ at 77 K. Quasi-periodic individual polaron motion next to the vertex in the domain structure. The data processing and visualisation correspond to Supplementary Video S2, with additional colour scale adjustment. The time counter represents the data time.

**Supplementary Video S4.**

https://www.youtube.com/watch?v=NnLMxC-J0k4&list=PLBvdEMM5_r8s&index=4

Sequence of Wigner-Seitz tessellations of the CDW on the surface of 1T-$TaS_2$ at 77 K. The Voronoi reconstruction is performed using standard methods for individual frames of the raw data (Supplementary Video S1). The colour scale is chosen to highlight non-hexagonal elements in the structure, and is shown in the video. The time counter represents the data time.

**Supplementary Video S5.**

https://www.youtube.com/watch?v=V6aMH6inCIs&list=PLBvdEMM5_r8s&index=5

Trajectories of polarons on the surface of 1T-$TaS_2$ at 77 K. Each element trajectory segment represents the shift between the current polaron position and the position of this polaron in the previous frame. The time counter represents the data time.

**Supplementary Video S6.**

https://www.youtube.com/watch?v=iUPI0FQTCb8&list=PLBvdEMM5_r8s&index=6

Temporal fluctuations for the Y-junction mode. A model simulation showing $\Delta 2(\boldsymbol{r},t)$ (blue) and a vector field proportional to the gradient $\nabla\phi(\boldsymbol{r},t)$ for the low-energy branch of the hybridised Higgs-Goldstone mode for the Y-junction mode.

**Supplementary Video S7.**

https://www.youtube.com/watch?v=5Neafefmvnw&list=PLBvdEMM5_r8s&index=7

Temporal fluctuations for circular (vortex-like) mode. A model simulation showing $\Delta 2(\boldsymbol{r},t)$ (blue) and a vector field proportional to the gradient $\nabla\phi(\boldsymbol{r},t)$ for the low-energy branch of the hybridised Higgs-Goldstone mode for the circular (vertex-like) mode.

**Supplementary Video S8.**

https://www.youtube.com/watch?v=4eQdD9BVlZ4&list=PLBvdEMM5_r8s&index=8

Simulation of spontaneous polaron self-organisation at 1/13 filling with non-trivial boundary conditions. The Monte Carlo simulations are performed as described in the main text. The data is represented in terms of the Wigner-Seitz tessellation for polarons. The black solid dots represent positions of the polarons, and the grey dots represent the atomic lattice. The colour scale is chosen to highlight non-hexagonal elements in the structure, and is shown in the video. The counter represents different "frames" in calculations.

**Supplementary Video S9.**

https://www.youtube.com/watch?v=GBw5AnIU36Y&list=PLBvdEMM5_r8s&index=9

Trajectories of polarons during self-organisation at 1/13 filling with non-trivial boundary conditions. The Monte Carlo simulations are performed as described in the main text. Each segment represents a single motion step of a single polaron. The counter represents frames in the calculations.